\documentclass[iop]{emulateapj}
\pdfoutput=1

\usepackage{natbib}
\usepackage{color}
\usepackage{bm}
\usepackage{hyperref}
\usepackage{float}
\usepackage{latexsym}
\usepackage{amsmath}
\usepackage{amssymb}
\usepackage{multirow}
\usepackage{rotate}
\usepackage{epstopdf}
\usepackage{graphicx}
\usepackage{url}

\include{epsf}
\DeclareGraphicsExtensions{.eps}
\slugcomment{Draft version v3.0, \today}

\begin{document}

%% definitions
\def \sb {mag\,arcsec$^{-2}$}
\def \mue {$<\!\mu\!>_{e}$}
\def \ellip {$\overline{E}$}
\def \triax {$\overline{T}$}
\def \sigellip {$\sigma_{E}$}
\def \sigtriax {$\sigma_{T}$}
\def \galfit {{\sc galfit}}
\def \ellipse {{\sc ellipse}}
\def \iraf {{\sc iraf}}
\def \emcee {{\sc emcee}}
\def \msun {M$_{\odot}$}
\def \atlas {ATLAS$^{3\text{D}}$}

\newcommand{\rsj}[1] {\textcolor{black}{(rsj: #1)}}
\newcommand{\new}[1] {\textcolor{black}{#1}}
\newcommand{\vect}[1]{\boldsymbol{\mathbf{#1}}}

\newcommand{\Engvs}{0.43}
\newcommand{\Eungvs}{+0.02}
\newcommand{\Elngvs}{-0.02}

\newcommand{\SEngvs}{0.11}
\newcommand{\SEungvs}{+0.02}
\newcommand{\SElngvs}{-0.01}

\newcommand{\Tngvs}{0.16}
\newcommand{\Tungvs}{+0.07}
\newcommand{\Tlngvs}{-0.06}

\newcommand{\STngvs}{0.04}
\newcommand{\STungvs}{+0.12}
\newcommand{\STlngvs}{-0.03}

\newcommand{\Elg}{0.51}
\newcommand{\Eulg}{+0.07}
\newcommand{\Ellg}{-0.06}

\newcommand{\SElg}{0.12}
\newcommand{\SEulg}{+0.06}
\newcommand{\SEllg}{-0.05}

\newcommand{\Tlg}{0.55}
\newcommand{\Tulg}{+0.21}
\newcommand{\Tllg}{-0.22}

\newcommand{\STlg}{0.06}
\newcommand{\STulg}{+0.17}
\newcommand{\STllg}{-0.05}


\title{The Next Generation Virgo Cluster Survey. VII. The intrinsic shapes of low-luminosity galaxies in the core of the Virgo cluster, and a comparison with the Local Group}
%\subtitle{}

\shorttitle{The shapes of low-mass galaxies in Virgo}
\shortauthors{S\'anchez-Janssen et al.}


\author{
Rub\'en S\'anchez-Janssen\altaffilmark{\ref{nrc}}, %
Laura Ferrarese\altaffilmark{\ref{nrc}}, %
Lauren A. MacArthur\altaffilmark{\ref{nrc},\ref{princeton}}, %
Patrick C\^ot\'e\altaffilmark{\ref{nrc}}, % 
John P. Blakeslee\altaffilmark{\ref{nrc}}, % 
Jean-Charles Cuillandre\altaffilmark{\ref{cfht}}, % 
Pierre-Alain Duc\altaffilmark{\ref{cea}}, % 
Patrick Durrell\altaffilmark{\ref{ohio}}, % 
Stephen Gwyn\altaffilmark{\ref{nrc}}, % 
Alan W. McConnacchie\altaffilmark{\ref{nrc}}, % 
Alessandro Boselli\altaffilmark{\ref{lam}}, % 
St\'ephane Courteau\altaffilmark{\ref{queens}}, % 
Eric Emsellem\altaffilmark{\ref{eso},\ref{lyon}}, % 
Simona Mei\altaffilmark{\ref{gepi},\ref{paris},\ref{caltech}}, % 
Eric Peng\altaffilmark{\ref{peking},\ref{kavli}}, % 
Thomas H. Puzia\altaffilmark{\ref{puc}}, % 
Joel Roediger\altaffilmark{\ref{nrc}}, % 
Luc Simard\altaffilmark{\ref{nrc}}, %
Fred Boyer\altaffilmark{\ref{ohio}}, %
Matthew Santos\altaffilmark{\ref{nrc}} % 
}

\email{ruben.sanchez-janssen@nrc-cnrc.gc.ca}

\newcounter{address}
\setcounter{address}{1}
\altaffiltext{\theaddress}{\refstepcounter{address}\label{nrc}NRC Herzberg Astronomy and Astrophysics, 5071 West Saanich Road, Victoria, BC, V9E 2E7, Canada}
%\altaffiltext{\theaddress}{\refstepcounter{address}\label{email} To whom correspondence should be addressed: \texttt{danfm@nyu.edu}}
\altaffiltext{\theaddress}{\refstepcounter{address}\label{princeton} Department of Astrophysical Sciences, Princeton University, Princeton, NJ 08544, USA}
\altaffiltext{\theaddress}{\refstepcounter{address}\label{cfht} Canada--France--Hawaii Telescope Corporation, Kamuela, HI 96743, USA}
\altaffiltext{\theaddress}{\refstepcounter{address}\label{cea} AIM Paris Saclay, CNRS/INSU, CEA/Irfu, Universit\'e Paris Diderot, Orme des Merisiers, F-91191 Gif sur Yvette cedex, France}
\altaffiltext{\theaddress}{\refstepcounter{address}\label{ohio} Department of Physics and Astronomy, Youngstown State University, One University Plaza, Youngstown, OH 44555, USA}
\altaffiltext{\theaddress}{\refstepcounter{address}\label{lam} Laboratoire d'Astrophysique de Marseille - LAM, Universit\'e d'Aix-Marseille \& CNRS, UMR7326, 38 rue F. Joliot-Curie, 13388, Marseille Cedex 13, France}
\altaffiltext{\theaddress}{\refstepcounter{address}\label{queens} Queen's University, Department of Physics, Engineering Physics and Astronomy, Kingston, Ontario, Canada}
\altaffiltext{\theaddress}{\refstepcounter{address}\label{eso} European Southern Observatory, Karl-Schwarzschild-Str. 2, 85748 Garching, Germany}
\altaffiltext{\theaddress}{\refstepcounter{address}\label{lyon} Centre de Recherche Astrophysique de Lyon and \'Ecole Normale Sup\'erieure de Lyon, Observatoire de Lyon, Université Lyon 1, 9 avenue Charles André, F-69230 Saint-Genis Laval, France}
\altaffiltext{\theaddress}{\refstepcounter{address}\label{gepi} GEPI, Observatoire de Paris, CNRS, University of Paris 
Diderot, Paris Sciences et Lettres (PSL), 61, Avenue de l'Observatoire 75014, Paris  France}
\altaffiltext{\theaddress}{\refstepcounter{address}\label{paris} University of Paris Denis Diderot, University of Paris Sorbonne Cit\'e (PSC), 75205 Paris Cedex
  13, France}
\altaffiltext{\theaddress}{\refstepcounter{address}\label{caltech} California Institute of Technology, Pasadena, CA 91125, USA}
\altaffiltext{\theaddress}{\refstepcounter{address}\label{peking} Department of Astronomy, Peking University, Beijing 100871, China}
\altaffiltext{\theaddress}{\refstepcounter{address}\label{kavli} Kavli Institute for Astronomy and Astrophysics, Peking University,
Beijing 100871, China}
\altaffiltext{\theaddress}{\refstepcounter{address}\label{puc} Institute of Astrophysics, Pontificia Universidad Cat\'olica de Chile, Av. Vicu\~na Mackenna 4860, 7820436 Macul, Santiago, Chile}

\begin{abstract}
%Low-mass galaxies dominate by number the galaxy luminosity function in all environments. 
%In order to put constraints on the variety of internal and external mechanisms driving the formation and evolution of dwarfs, it is both necessary and challenging to robustly characterize these faint systems observationally.
\new{We investigate the intrinsic shapes of low-luminosity galaxies in the central 300 kpc of the Virgo cluster using deep imaging obtained as part of the Next Generation Virgo Cluster Survey (NGVS).
We build a sample of nearly 300 red-sequence cluster members in the yet unexplored $-14 < M_{g} < -8$ magnitude range, and we measure their apparent axis ratios, $q$, through S\'ersic fits to their two-dimensional light distribution--which is well described by a constant ellipticity parameter.
The resulting distribution of apparent axis ratios is then fit by families of triaxial models with normally-distributed intrinsic ellipticities, $E = 1-C/A$, and triaxialities,  $T = (A^{2} - B^{2})/(A^{2}-C^{2})$. 
We develop a Bayesian framework to explore the posterior distribution of the model parameters, which allows us to work directly on discrete data, and to account for individual, surface brightness-dependent axis ratio uncertainties. 
For this population we infer a mean intrinsic ellipticity \ellip $=\Engvs^{\Eungvs}_{\Elngvs}$, and a mean triaxiality \triax $= \Tngvs^{\Tungvs}_{\Tlngvs}$.
This implies that faint Virgo galaxies are best described as a family of thick, nearly oblate spheroids with mean intrinsic axis ratios $1:0.94:0.57$.
The core of Virgo lacks highly elongated low-luminosity galaxies, with 95 per cent of the population having $q > 0.45$.
We additionally attempt a study of the intrinsic shapes of Local Group (LG) satellites of similar luminosities.
For the LG population we infer a slightly larger mean intrinsic ellipticity \ellip $=\Elg^{\Eulg}_{\Ellg}$, and the paucity of objects with round apparent shapes translates into more triaxial mean shapes,  $1:0.76:0.49$.
Numerical studies that follow the tidal evolution of satellites within LG-sized halos are in good agreement with the inferred shape distributions, but the mismatch for faint galaxies in Virgo highlights the need for more adequate simulations of this population in the cluster environment.
We finally compare the intrinsic shapes of NGVS low-mass galaxies with samples of more massive quiescent systems, and with field, star-forming galaxies of similar luminosities.
We find that the intrinsic flattening in this low-luminosity regime is almost independent of the environment in which the galaxy resides--but there is a hint that objects  may be slightly rounder in denser environments.
The comparable flattening distributions of low-luminosity galaxies that have experienced very different degrees of environmental effects suggests that internal processes are the main drivers of galaxy structure at low masses--with external mechanisms playing a secondary role.}\\
\end{abstract}





%%%%%%%%%%%%%%%%%%%%%%%%%%
\section{Introduction}

%Low-mass galaxies dominate by number the galaxy luminosity function in all environments. 
The faint end of the satellite luminosity function is dominated by systems with smooth appearance, low surface brightness, and little gas content or recent star formation activity--the so-called dwarf spheroidal (dSph) galaxies \citep[][and references therein]{Mateo1998,vandenBergh2000,Grebel2003,Tolstoy2009}. 
In the Local Group, their luminosities and stellar velocity dispersions are similar to those of globular clusters, but their much larger half-light radii imply substantial total masses, making these faint galaxies the most dark matter dominated systems known in the Universe \citep[e.g.,][]{Walker2009}.

Two observed properties of this population of faint satellites are known to place strong constraints on $\Lambda$CDM models of galaxy formation.
First, their number is short by one to two orders of magnitude compared to the expectations from the mass function of cold dark matter halos (\citealt{Klypin1999,Moore1999}; but see \citealt{Hargis2014}). 
Attempts to resolve this discrepancy traditionally invoke baryonic mechanisms that lower the efficiency of, or even suppress, star formation at these mass scales--including high gas cooling times, reionization, stellar feedback, and early environmental effects \citep{White1978,Dekel1986,Bullock2000,Benson2003,Mayer2007,Okamoto2008}.
 In passing, we note that the abundance of the gas-rich, star-forming counterparts which exist in the same luminosity regime in the field also seems to be lower than what is predicted by  $\Lambda$CDM \citep{Klypin2015}. Obviously, environmental effects can not be invoked in this case.
Second, the inferred central dark matter densities in these systems favor constant density cores instead of the steeper density profiles found in dark matter simulations \citep{Walker2011}. 
Hydrodynamic simulations suggest that the removal of these dark matter cusps occurs through energetic outflows from supernovae, a mechanism that requires frequent and centrally concentrated bursts of star formation \citep{Governato2010,Pontzen2012,Teyssier2013}.
 
\citet{Penarrubia2012} point out that these two  effects place competing requirements for the star formation efficiency in low-mass halos, creating tension in CDM models of galaxy formation. 
This friction can be alleviated, however, by a combination of baryonic physics and (enhanced) tidal stripping that bring both the inner mass profiles and the abundance of satellites in simulations in agreement with the observations \citep{Zolotov2012,Brooks2013}.
Indeed, low-mass quiescent satellites tend to be found in the vicinity of luminous ($L \gtrsim L^{*}$) central galaxies \citep{Geha2012,rsj2013b}, and dynamical processes have long been proposed to influence their evolution \citep{Penarrubia2008,Kazantzidis2011}.
 
In order to put  constraints on the plethora of internal and external mechanisms potentially shaping the properties of faint satellites, it is mandatory to robustly characterize these systems observationally. 
In this sense, it is only natural to start with the most accessible observables, i.e., their shapes and structural parameters. 
Importantly, the relatively shallow potential wells in which these galaxies reside ensure that both photoionization and stellar feedback processes, as well as environmental effects, leave recognizable imprints on their stellar structure.
For instance, \citet{Kaufmann2007} show that baryonic processes that act to pressurize gas within low-mass halos naturally lead to the formation of thicker, puffier galaxies towards lower luminosities as a result of their progressively diminishing angular momentum support. 
In particular, feedback from bursty star formation can result in strong potential fluctuations caused by gas outflows and inflows. The ensuing heating of stellar orbits can drive significant structural variation, including outward migration and thickening of the stellar body \citep{Teyssier2013,El-Badry2015}. 
Other sources of internal  heating, such as dynamical interactions with dark subhalos, may play a role as well \citep{Helmi2012,Starkenburg2015}.
Alternatively, tidal effects have the potential to significantly modify the structural (and dynamical) properties of disky systems, creating pressure-supported triaxial objects that in many aspects resemble the satellites of cluster- and group-sized halos \citep{Mastropietro2005,Smith2010,Kazantzidis2011,Lokas2012,Bialas2015}.
\new{Gas stripping mechanisms have a smaller direct impact on the stellar structure of low-mass systems, except for extreme cases where the gas component represents a large fraction of the total mass budget \citep[see][]{Smith2013a}. They can nevertheless contribute to the thickening of the stellar body on timescales of a few Gyr by cutting the supply of young stars with low velocity dispersion, followed by subsequent diffusion through two-body relaxation.}

Until recently, the intrinsic shapes of faint quiescent galaxies have not been studied extensively. This  stems from the fact that these are intrinsically faint objects, and the great majority of known and well-characterized dSphs are Local Group satellites--and the number statistics there are low. For example, the complete sample of Milky Way (MW) dSphs studied by \citet{Martin2008} includes only 21 galaxies, 11 of which are fainter than $M_{V} = -7.5$ mag. Interestingly, while the classical MW satellites show small apparent ellipticities, the faint population display more elongated shapes. This immediately prompts the suspicion that tidal interactions with the MW are behind the origin of this highly elongated galaxies. 
Along these lines, \citet{Lokas2012} find good general agreement between the apparent shapes of MW satellites and simulations of tidally stirred disks orbiting an MW-sized halo.
\new{Recently, \citet{Salomon2015} have inferred the intrinsic ellipticity of M31 satellites under the assumption that their visible components have prolate shapes, and after exploring different possibilities for the orientation of the major axis. They find that the population has a mean intrinsic axis ratio $\approx 0.5$, but several satellites are consistent with being intrinsically more elongated.}
%But these two works limit their analysis to the apparent ellipticities of MW dSphs, and do not discuss their intrinsic shapes in detail.
Because the orientation of individual galaxies in the sky is hard to obtain, intrinsic shapes are usually derived on statistical grounds using relatively large samples. 
Low number statistics still make this task challenging in the Local Group (but again, see \citealt{Salomon2015}).
It is however now possible  to study large numbers of faint, quiescent galaxies in the nearby Virgo cluster thanks to the Next Generation Virgo Cluster Survey (NGVS).
%In order to investigate the intrinsic shape of dSphs, much larger samples of at least hundreds of objects are required. 
%while low number statistics make this task difficult in the local group...
%A task almost  insurmountable in the Local Group, it is now possible in the nearby Virgo cluster thanks to the Next Generation Virgo Cluster Survey (NGVS). 

The NGVS \citep[][F12 hereafter]{Ferrarese2012} is a Canada-France-Hawaii Telescope (CFHT) optical imaging survey of the Virgo cluster covering 104 deg$^2$, from its core to the virial radius. The MegaCam deep imaging delivers a 2\,$\sigma$ surface brightness limit $\mu_{g} \approx 29$ \sb, with excellent image quality (median FWHM = 0.54 arcsec in the $i$-band). 
This has allowed us to detect and characterize, in the core region alone ($R_{p} \lesssim 300$ kpc), hundreds of previously unknown low-mass galaxies down to the luminosity limits of the classical MW satellites, $M_{V} \sim -8$ mag. This constitutes a major breakthrough, not only because of the order(s) of magnitude increase in the sample size, but also because Virgo represents a completely different environment from the Local Group.
Here we use the observed apparent axis ratios of a sample of nearly 300 low mass galaxies in the core of the Virgo cluster to infer the distribution of their intrinsic shapes. We will show that these low-mass Virgo satellites are a family of thick, nearly oblate spheroids.
%Other papers in the NGVS series related to faint stellar systems in Virgo include the structural and dynamical properties of ultra-compact dwarfs \citep[][Liu et al. 2015, submitted]{Zhang2015}; the cluster-wide population of globular clusters \citep{Durrell2014}; the stellar populations and kinematics of compact, low-mass galaxies in Virgo \citep{Guerou2015}; the discovery of tidally disrupted satellites around an $L^{*}$ disc galaxy \citep{Paudel2013}; the physical classification of stellar and galactic sources using optical and NIR imaging \citep{Munoz2014}; and the determination of photometric redshifts for background sources \citep{e.g., raichoor2014}.

This paper is organized as follows. In Section\,\ref{sect:data} we present the dataset, and the detection and structural characterization of faint Virgo galaxies. We then describe the Bayesian framework developed to infer their intrinsic shapes in Section\,\ref{sect:method}, followed by the main results of our analysis in Section\,\ref{sect:ngvs_shapes}. In Section\,\ref{sect:discuss} we discuss our results, and compare them with previous efforts in the literature. Finally, in Section\,\ref{sect:summary} we summarize our findings.

Throughout this work we adopt the convention that apparent (i.e., observed) quantities are represented with lower case symbols, whereas upper case symbols are used for intrinsic parameters. Thus, a triaxial galaxy with intrinsic semi axes $A \ge B \ge C$ has an apparent ellipticity ($\epsilon$) and an apparent axis ratio ($q$) related through $\epsilon = 1- q$.
We furthermore use a common distance modulus $(m-M)=31.09$ mag for all candidate Virgo members, corresponding to the mean distance of $D=16.5$ Mpc to the Virgo cluster derived through the surface brightness fluctuations method \citep{Mei2007,Blakeslee2009}.

 %%%%%%%%%%%%%%%%%%%%%%%%%%

%%%%%%%%%%%%%%%%%%%%%%%%%%
\section{Detection and characterization of low-mass quiescent galaxies in Virgo}
\label{sect:data}

\subsection{Sample selection in the NGVS}
In this work we use data from the NGVS. Here we only provide a brief overview of the most relevant characteristics of the survey, but  the interested reader is referred to F12 for a more detailed description of the survey, datasets and science goals.
The NGVS is a CFHT/MegaCam imaging survey of the Virgo cluster covering 104 deg$^{2}$ in the $u^{*}giz$ bandpasses, supplemented with $K_{s}$-band imaging for a fraction of the area \citep{Munoz2014}, as well as a series of dedicated spectroscopic follow-ups.
The central 4 deg$^2$ around M87 ($R \lesssim 300$ kpc) were observed in a Pilot Programme (PP) that included the $r$-band as well, and all objects in this study come from this core region. 
By design, the survey has an excellent image quality, with a  median seeing FWHM $\approx 0.8$ arcsec in the $g$-band (corresponding to $\sim65$ pc at the distance of Virgo).
The \emph{Elixir-LSB} reduction pipeline \citep{Ferrarese2012,Duc2015} delivers a superb flattening of the sky background, with peak-to-peak residuals of $\sim0.2$ per cent that result in a 2\,$\sigma$ surface brightness limit $\mu_{g} \approx 29$ \sb\ for the final stacked frames.
The NGVS is therefore perfectly positioned to carry out, for the first time, an investigation of the properties of low-mass satellites in the Virgo cluster.
The identification and characterization of Virgo cluster members will be 
discussed in detail in an upcoming paper of this series (Ferrarese et 
al., submitted); here we summarize the most relevant aspects. 

The detection of extended, low surface brightness Virgo galaxies in a field 
contaminated with a far larger number of foreground stars and background 
objects is a challenge; conventional codes, such as SExtractor \citep{Bertin1996}, tend to regard low surface brightness objects as 
belonging to compact and/or brighter contaminants. To circumvent the 
problem, a ring median filter \citep{Secker1995} was first applied to each 
$g$-band image (the NGVS images with the highest signal-to-noise ratio), 
with radius adjusted to suppress all unresolved sources (stars and 
globular clusters), as well as compact background galaxies. A specific 
optimization of SExtractor was then run on the median-kernel smoothed images, thus 
enabling the identification of low surface brightness sources that are potential 
Virgo members. 

\new{Our low-mass galaxy sample explicitly excludes the class of so-called ultra-compact dwarfs \citep[UCDs;][]{Hilker1999,Drinkwater2000}. These objects have sizes and luminosities bridging the main locii occupied by low-mass galaxies and globular clusters in structural scaling relations. As a result, the nature of UCDs is still hotly debated in the literature, and they probably encompass objects of both galactic and star cluster origin \citep[e.g.,][]{Hacsegan2005,Mieske2008,Chiboucas2011,Penny2014,Norris2014,Seth2014}. We refer the interested reader to \citet{Durrell2014} for an overview of the spatial distribution of bright compact stellar systems in Virgo; to \citet{Zhang2015}, where we have published a kinematical analysis of $\sim100$ spectroscopically confirmed UCDs in the same Virgo core area we study here; and to \citet{Liu2015} for an investigation of their photometric properties.}

The resulting galaxy catalogue does of course not 
discriminate between Virgo members and contaminants, and therefore the next 
challenge is to define membership criteria. These are based on the 
location of each galaxy in a multi-parameter space defined by a 
combination of i) galaxy structural parameters, specifically size and 
surface brightness, measured in 
appropriately masked original images using \galfit\ \citep{Peng2002}; ii) photometric redshifts,\footnote[1]{We note here that photometric redshifts are mainly useful to identify and discard obvious background systems rather than selecting Virgo members, as their typical accuracy at low redshift, $\delta z/(1+z) \sim 0.05$, is not good enough to establish secure cluster membership \citep[e.g.,][]{raichoor2014}.} based on 
$u^{*},g,i,z$, and, when available, $r$-band photometry; and iii) an index 
measuring the strength of residual structures in images created by 
subtracting from each galaxy the best fitting \galfit\ model. The exact 
combination of axes in this space was selected to allow for maximum 
separation of known Virgo and background sources. The former comprise 
spectroscopically confirmed (mostly) VCC galaxies, while the latter are 
identified, using the same procedure described above, in four control 
fields located three virial radii away from M87, and presumed to be 
devoid of cluster members.

The reliability of the procedure was tested in two independent ways. 
First, 36,500 artificial galaxies were injected in the 
frames spanning the full range of luminosity and structural parameters 
expected for genuine Virgo galaxies, and then processed as described 
above. With few exceptions (for instance galaxies that land in the 
immediate vicinity of bright saturated stars or  near the cores of high 
surface brightness galaxies), all galaxies with surface brightness high 
enough to be visible in the frames were indeed recovered. A detailed 
discussion of the completeness of the data and the biases in the 
recovered parameters is included in Ferrarese et al. (submitted). 
Second, four NGVS team members (LF, PAD, PD, and FB) independently inspected the central four square 
degrees of the PP region and identified, by eye, all objects 
that appeared to be bona-fide Virgo members; again with few exceptions 
all such galaxies were detected and correctly identified as bona-fide 
Virgo members by the code.
We stress here that the proximity of the Virgo cluster, together with the remarkable tightness of galaxy scaling relations even at these faint luminosities, are key in providing a clean separation between cluster members and background galaxies even in the absence of spectroscopic information \citep[see also][]{Michard2008,Rines2008}.

\begin{figure}[!t]
\includegraphics[angle=0,width=0.5\textwidth]{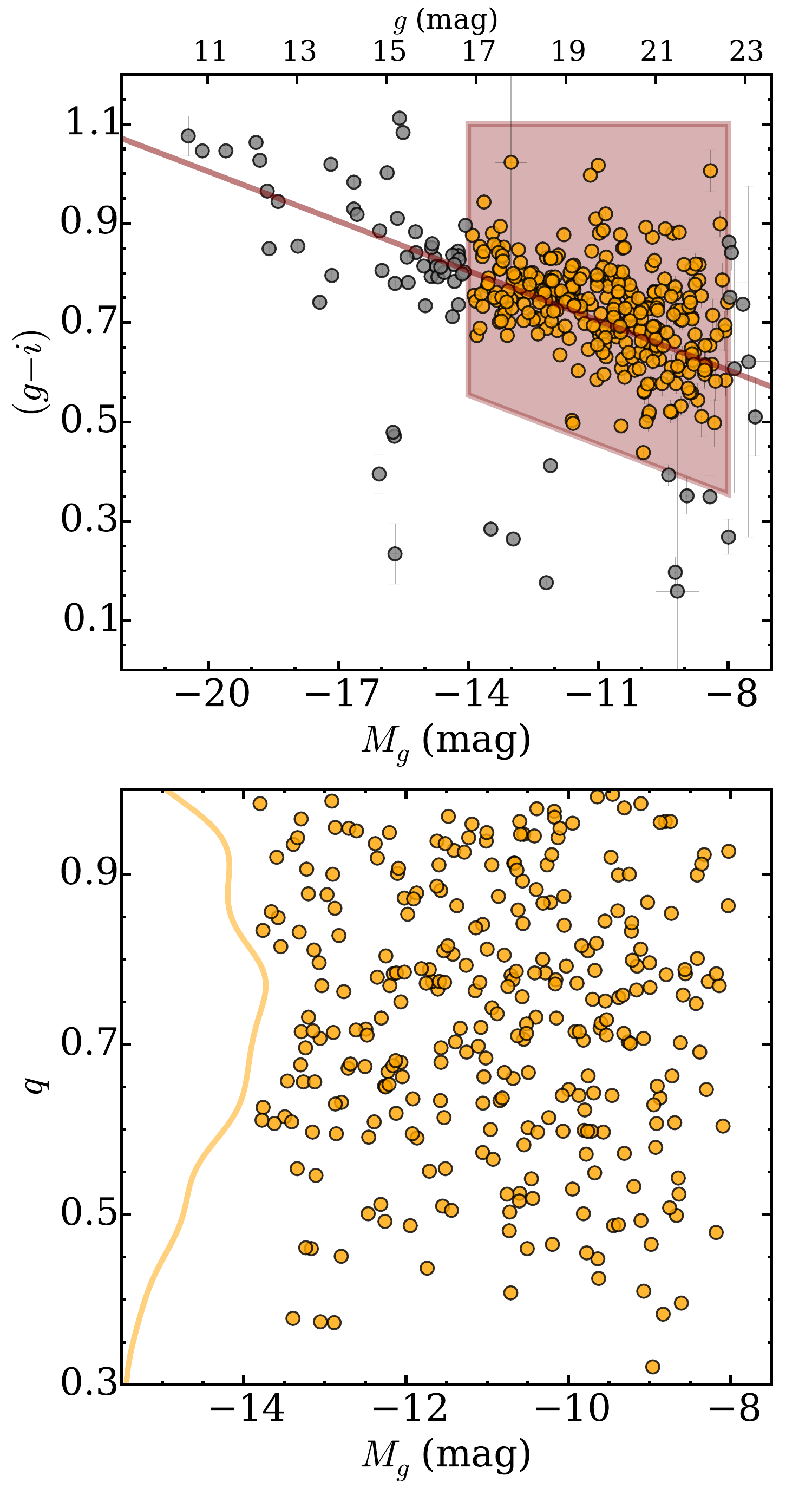}
\caption{\emph{Top:} ($g-i$) color-magnitude diagram of galaxies in the core of the Virgo cluster. Low-luminosity Virgo candidates are selected as galaxies having absolute magnitudes in the $-14 \leq M_{g} \leq -8$ mag range, and colors consistent with being red-sequence cluster members--as indicated by the shaded region. 
\emph{Bottom}: apparent axis ratios for the low-luminosity sample as a function of absolute magnitude. The solid line is a simple gaussian kde estimate of the $q$ distribution. Note the absence of very elongated galaxies, and the lack of dependence on luminosity.}
\label{fig:cmd}
\end{figure}

\begin{figure}[!t]
\includegraphics[angle=0,width=.5\textwidth]{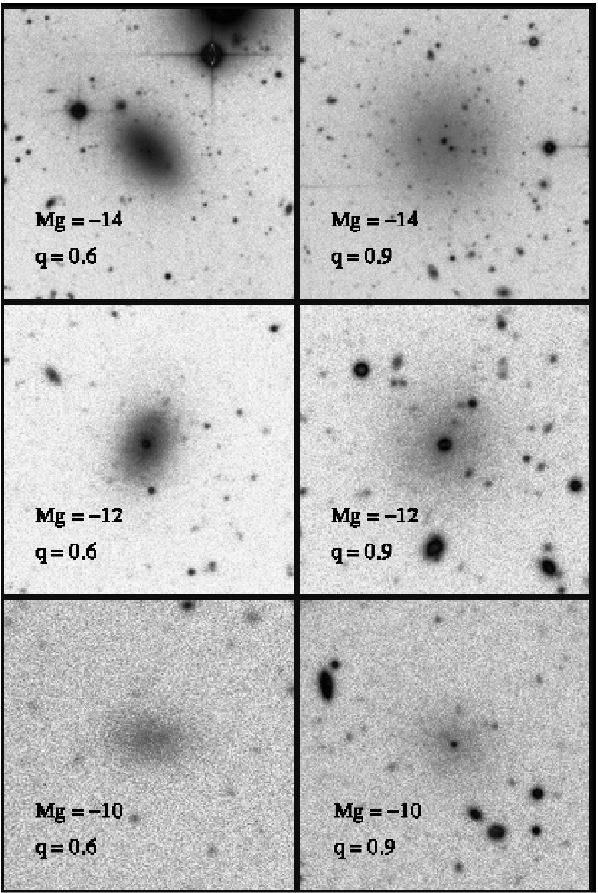}
\caption{NGVS negative postage stamps of six representative faint Virgo galaxies. Rows correspond to galaxies with similar absolute magnitudes, while systems with similar apparent axis ratios are shown along columns. The left column displays objects with significant apparent elongation ($q \approx 0.6$), while nearly round ($q \approx 0.9$) objects are shown on the right column.}
\label{fig:pstamps}
\end{figure}

Figure\,\ref{fig:cmd} (top) presents the $(g-i)$ color-magnitude diagram of all 404 candidate Virgo galaxies in the PP area (except M87). As expected given that we cover the inner regions of the cluster, the great majority of objects fall on a well-defined red sequence all the way down to the faintest luminosities probed \citep[see also][]{rsj2008,Boselli2014a}.
In this work we focus on the shapes of faint, low surface brightness, quiescent galaxies, analogous to the dSph satellites in the LG. 
It is now well established that this label encompasses objects with quite varied star formation histories \citep[e.g.,][]{Weisz2014a}. Throughout this work we keep this terminology when referring to LG satellites for historic reasons, but we will avoid it as much as possible when referring to faint, quiescent Virgo galaxies.
Because detailed star formation histories are not available yet for Virgo galaxies of these low masses, we construct our sample of quiescent objects from galaxies with colors consistent with being red-sequence cluster members (see the shaded region in the top panel of Fig.\,\ref{fig:cmd}). The lower boundary is set at 0.25 mag below the best-fit color-magnitude relation, whereas $(g-i)=1.1$ defines the red end of our selection function . Furthermore, we only select objects in the $-14 \leq M_{g} \leq -8$ mag range.\,\footnote[2]{This roughly corresponds to stellar masses $3\times10^{5} \lesssim \text{M$_{\star}$/M$_{\odot}$} \lesssim 7\times10^{7}$, for a characteristic stellar mass-to-light ratio $\text{M$_{\star}$}/L_{V} = 1.6~\text{M$_{\odot}$}/L_{\odot}$ \citep{Woo2008}.}
With this luminosity cut we guarantee our galaxies to be genuinely faint systems; we match the luminosity range of other low-luminosity samples in the LG and in the Local Universe that will serve as comparison; and we probe a yet unexplored galaxy mass range in Virgo--or any other cluster, for that matter.
 
We also explored a visual morphological classification with the aid of composite $giz$ color images and unsharp-masked $g$-band frames. Faint quiescent systems were defined as objects with no evident star formation activity and smooth, elliptical appearances. We found that both the color and morphological classifications produced very similar samples, and therefore our results do not strongly depend on the exact definition of what is a quiescent low-mass galaxy.
In the following analysis we use the color selection simply because it results in a more quantitative and reproducible selection function.   
%This sample comprises 316 red-sequence, low-luminosity galaxies in the core of the Virgo cluster.

\subsection{Measurements of apparent axis ratios}
\label{sect:axis_ratios}

Structural parameters for low-luminosity galaxies in the NGVS are obtained from both surface brightness profiles using the \iraf\ task \ellipse\ \citep{Jed1987}, and two-dimensional model fits using \galfit. 
The derived structural parameters include, but are not limited to, the effective radius and surface brightness, the S\'ersic shape parameter, the position angle, and the apparent axis ratio $q$--which is the most relevant quantity for this study.\,\footnote[3]{In this work we use the mean axis ratio derived by \ellipse\ in the $1 \leq r/r_e \leq 2$ interval as a point estimate for $q$.} 
Here we will only use structural parameters derived in the $g$-band, as it is the band with the highest signal-to-noise ratio in the NGVS. We have verified that our results do not change if we use either the $r$- or the $i$-band data. This is expected if these faint systems, like the dSphs in the LG, have relatively uniform stellar populations, and little obscuration caused by dust.
  
In both approaches the galaxy body is modeled with a \citet{Sersic1968} function, but only in the case of the isophotal analysis with \ellipse\ is the ellipticity allowed to vary with radius--\galfit\ uses constant axis ratio models.
If necessary, a central nuclear component is included in the fits. 
With typical half-light radii $r_{h} < 0.3$ arcsec at the distance of Virgo \citep{Cote2006,Turner2012}, most of these nuclei remain unresolved in the NGVS images. They are simply modelled as PSF components \citep[see also][]{denBrok2014}. %, and with a S\'ersic function in the one-dimensional case.
%This is potentially an advantage for objects with prominent stellar nuclei (see for example the galaxy in the middle panels of Fig.\,\ref{fig:pstamps}), because if the  central component is spatially resolved, it will tend to circularize the luminosity-weighted estimate of the axis ratio for a model with constant ellipticity.
%Nevertheless, even if we were not including a nuclear component in the fits, this should not represent a problem for most faint galaxies in the NGVS because i) the fraction of nucleated objects decreases towards fainter luminosities \citep{Cote2006, Lisker2006, Turner2012}; and ii) with typical half-light radii $r_{h} < 0.3$ arcsec, the majority of this nuclear population will remain unresolved in our images.
Finally, all the parameters derived from \galfit\ fits are seeing-deconvolved, while \ellipse\ measures seeing-convolved axis ratios. 
Once again, this should not impact the global structural parameters of objects that have shallow surface brightness profiles and are well resolved. For reference, the smallest effective radius for an $M_{g} \approx -8$ Virgo galaxy in our sample is $r_{e} \approx 2.5$ arcsec ($\sim200$ pc), which is more than a factor three larger than the $g$-band FWHM.

In this work we use the \galfit\ axis ratios for a practical reason. For a non-negligible fraction of the faintest galaxies, the \ellipse\ fits do not converge when all  parameters are allowed to vary due to a combination of small apparent ellipticity and very low central surface brightness. For these objects the fits were carried out with the axis ratio fixed at $q$ = 0.95. Using the  \ellipse\ parameters would create an artificial peak in the distribution of observed axis ratios at this value, and would prevent us from modeling it appropriately (see Sect.\,\ref{sect:method}).

The observed distribution of \galfit\ axis ratios for our faint Virgo galaxies is plotted as a function of $g$-band absolute magnitude in the bottom panel of Figure\,\ref{fig:cmd}.
It is interesting to note the paucity of low-luminosity systems in the core of Virgo displaying highly elongated shapes: 95 per cent of the population have $q \gtrsim 0.45$. This is an important result considering we are only sampling objects with projected distances $\lesssim 300$ kpc from M87, that is, the highest density regions of the cluster where tidal forces are maximized.
Additionally, and at least in this mass regime and cluster region, there does not seem to be any dependence of the apparent shapes $q$ on luminosity or clustercentric distance. This guarantees that we can use the entire sample to infer the intrinsic shape of the population.
Figure\,\ref{fig:pstamps} shows negative postage stamps of six representative galaxies in the PP region, illustrating the range of luminosities and apparent shapes of this faint population, and the quality of the NGVS imaging.

\subsubsection{Low-luminosity galaxies in Virgo have flat ellipticity profiles}
\label{sect:ellip_radial} 

\begin{figure}[!t]
\includegraphics[angle=0,width=.5\textwidth]{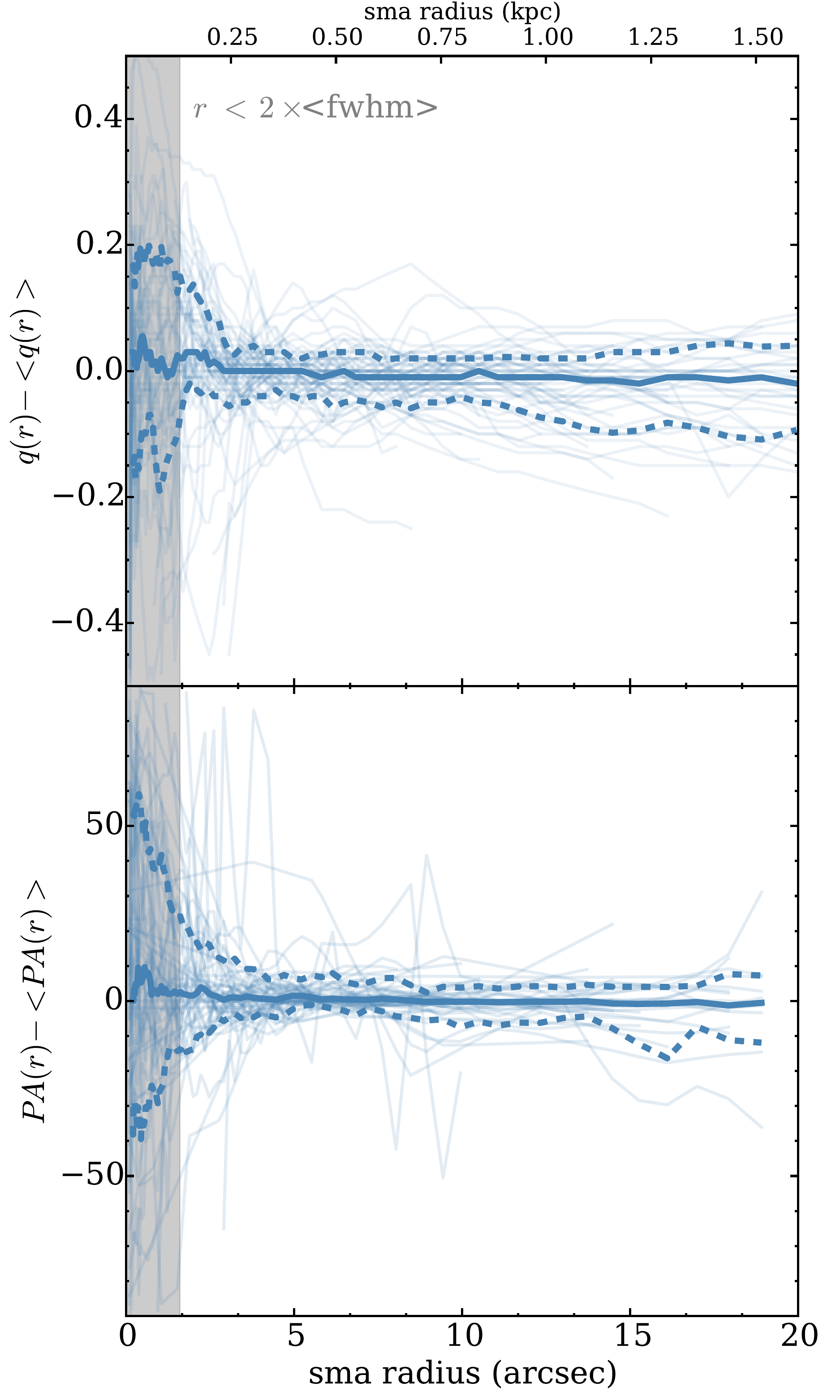}
\caption{\new{Radial variation about the mean of the isophotal axis ratio (top) and position angle (bottom) for low-luminosity Virgo galaxies. The light curves show the individual radial profiles, with their mean value subtracted. The solid thick line indicates the median of the variation at fixed radius, and the thick dashed lines enclose 68 per cent of all the curves. These low-luminosity systems exhibit almost flat axis ratio profiles, and no evidence for isophotal twisting.}}
\label{fig:ellip_radial}
\end{figure}

\new{One potential limitation of the \galfit\ surface-brightness-weighted $q$ estimates is that they do not necessarily represent the shape of the galaxy at any particular radius. This could introduce a bias if, for example, the ellipticity profile presented significant radial variation, or if the amplitude of that variation was luminosity-dependent.
We can rule out this is the case for our sample of NGVS low-mass objects from an analysis of the quality of the \galfit\ fits. S\'ersic models with constant ellipticity provide excellent fits to this population, with a median $\chi_{\nu}^{2} = 1.00$,  and a dispersion $\sigma_{\chi_{\nu}^{2}} = 0.11$. Moreover, fit residuals do not show any radial structure, thus ruling out strong isophotal twisting as well.}

\new{We also use the \ellipse\ fits to study the radial ellipticity profiles in our sample. Thin lines in Fig.\,\ref{fig:ellip_radial} show, for objects having \mue\ $<27$ \sb, the individual $q$  and PA radial profiles after their mean value has been subtracted. The solid thick line indicates the median of the variation at fixed radius, and the thick dashed lines enclose 68 per cent of all the curves. The profiles are remarkably flat, with little variation about the mean along a large radial range. We note that, because of this flatness, this behavior holds even if we scale the profiles by each individual effective radius. We however prefer this representation because it allows us to show the radial range that is potentially affected by the seeing PSF (shaded region in Fig.\,\ref{fig:ellip_radial}). The profiles show larger scatter at $r < 2\,\text{FWHM}$, with a slight preference for rounder shapes--consistent with mild circularization caused by seeing, as well as the presence of nuclear star clusters in some of these galaxies. 
In summary, low-luminosity quiescent Virgo galaxies are systems with remarkable isophotal simplicity and well described by a constant ellipticity parameter and major axis position angle. This lends support to the notion that homologous triaxial models are a good representation of their intrinsic shapes.}

\subsubsection{Accuracy of the axis ratio measurements}
\label{sect:reliability}
\begin{figure}[!t]
\includegraphics[angle=0,width=0.5\textwidth]{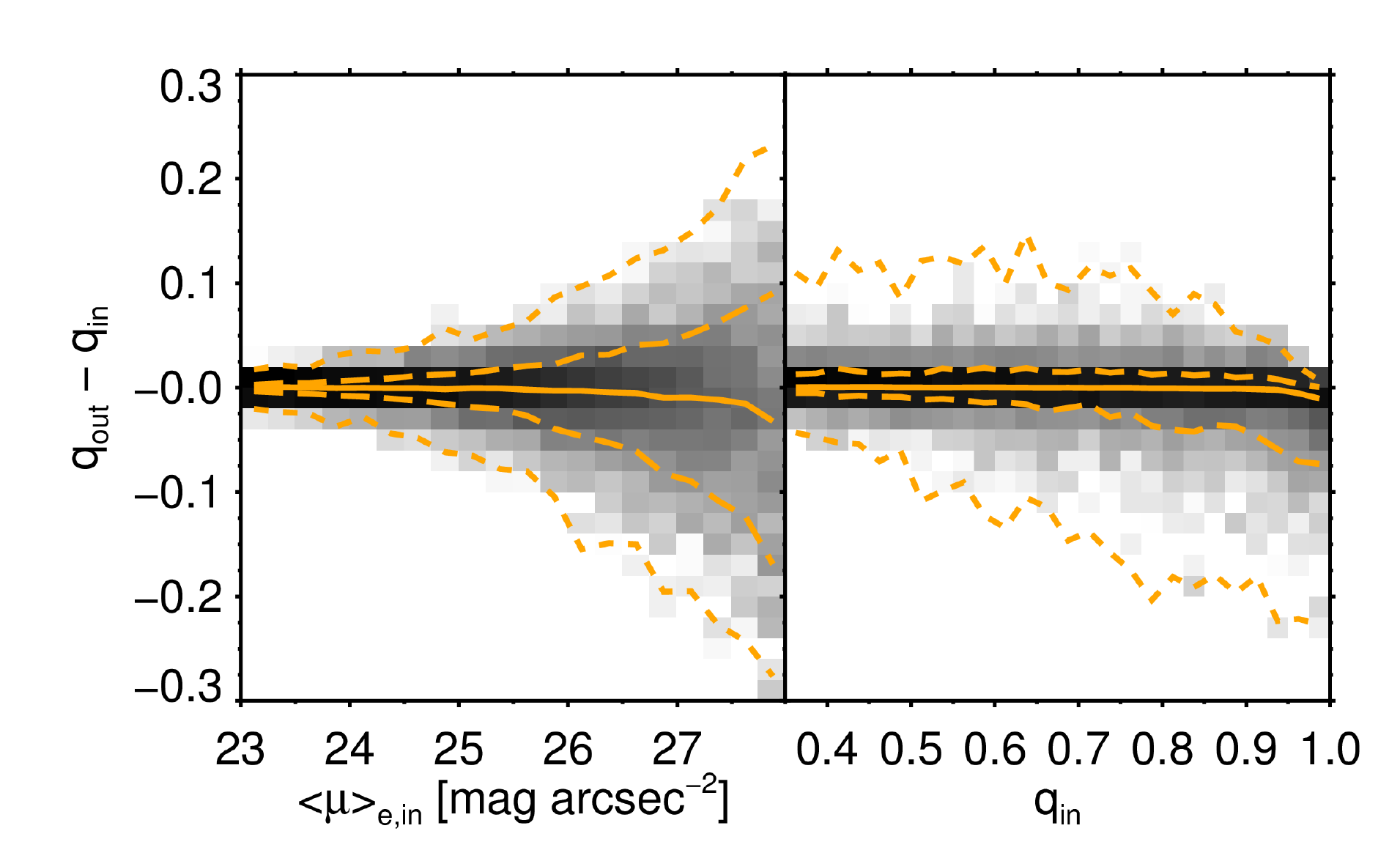}
\caption{Error in the recovered axis ratios of simulated single-S\'ersic galaxies as a function of their mean effective surface brightness (left panel), and axis ratio (right panel). The ordinates show the difference between the input axis ratios and the recovered \galfit\ values, with the grey scale showing the density of galaxies in bins of \mue\ and $q$. The long-dashed and short-dashed lines enclose 68 and 95 per cent of the data points, respectively.}
\label{fig:axrat_err}
\end{figure}

Assessing the reliability of the axis ratio measurements for galaxies of such low central surface brightness is of paramount importance for this study, and we have carried out several consistency checks.
First, as we have already mentioned, we simulated and added 36,500 single-S\'ersic model galaxies to the images, sampling a range in structural parameters. 
The galaxy models are added to real NGVS frames after being convolved with the appropriate PSF for that randomly selected image location, and the simulated galaxy is fitted with \galfit\ in the same way the real Virgo galaxies are. 
Figure\,\ref{fig:axrat_err} shows the difference between recovered and input axis ratios for the simulated galaxies as a function of their true mean effective surface brightness (left panel), and their true axis ratio (right panel). The grey scale shows the density of galaxies in bins of constant $<\!\mu\!>_{e,in}$ or $q_{in}$, with the solid line indicating the median offset. The long-dashed and short-dashed lines enclose 68 and 95 per cent of the data. It is clear that the axis ratios are very robustly determined down to mean effective surface brightnesses as low as $<\!\mu\!>_{e}\, \approx 28$ \sb, displaying no biases and a maximum 1\,$\sigma$ error $\approx 0.1$.
At fixed \mue\ the error distribution is reasonably well described by a normal function.
The right panel shows that the error distribution as a function of input axis ratio is highly peaked around zero. The  bias toward recovering smaller axis ratios for larger $q_{in}$ is only a natural consequence of the limitation that, by definition, $q \le 1$.
This is an often neglected but important bias, and in Sect.\,\ref{sect:bayes} we will discuss how to properly account for it in the analysis.

\begin{figure}[!t]
\includegraphics[angle=0,width=0.5\textwidth]{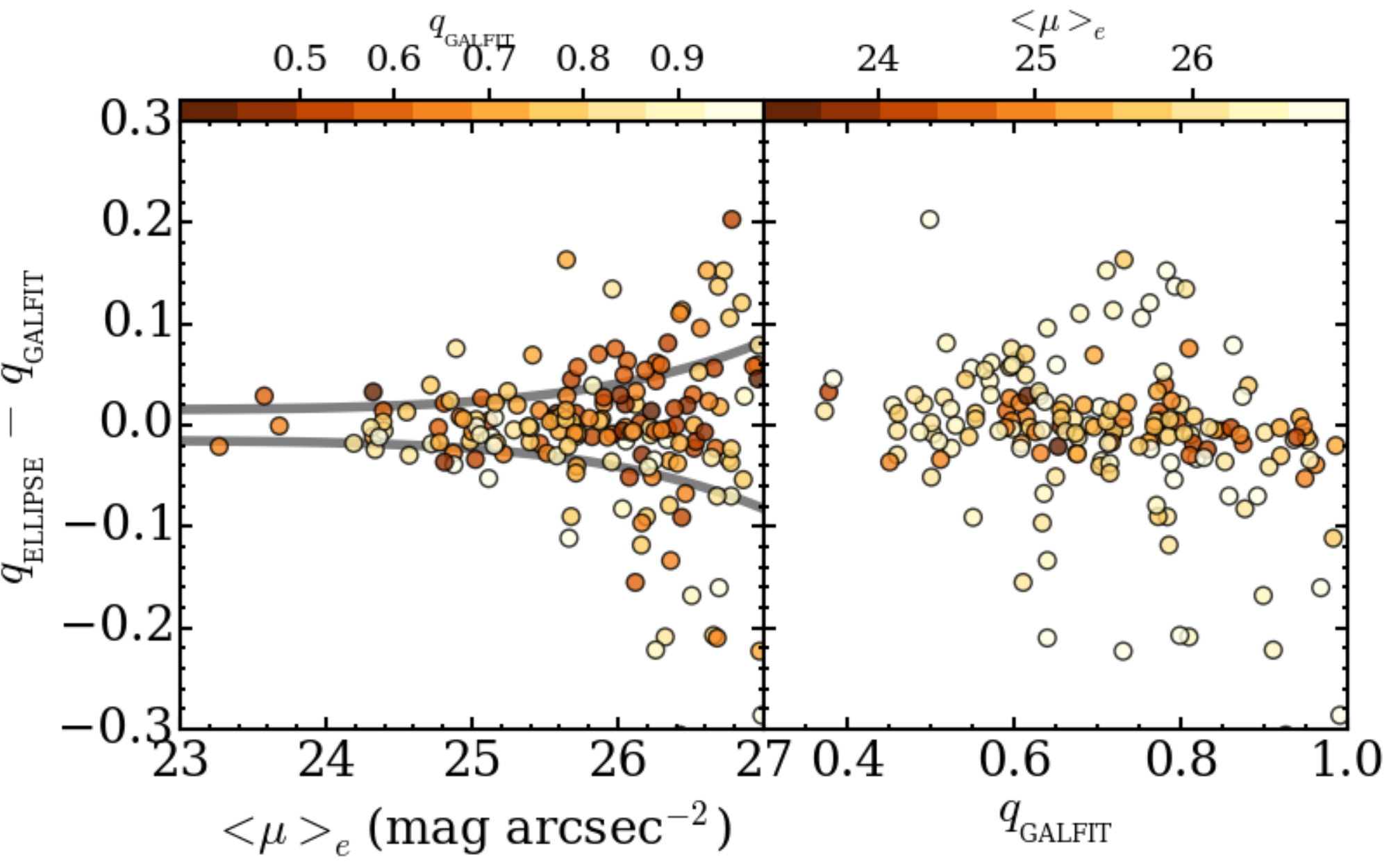}
\caption{\new{Difference between the mean axis ratios measured with \ellipse and the \galfit\ $q$ values for NGVS low-mass galaxies. The left panel shows differences as a function of their mean effective surface brightness, whereas the right panel shows them as a function of the \galfit\ axis ratio measurements. Symbols are color-coded according to their $q$ and \mue, respectively. The solid grey lines in the left panel indicate the $1\sigma$ \mue-dependent uncertainty we use in our analysis.}}
\label{fig:axrat_comparison}
\end{figure}

The uncertainties in $q$ derived from these simulations can be regarded as lower limits only, in the sense that both the PSF and the galaxy light profile are perfectly described by the fitting functions, and the ellipticity is radially constant by construction. 
To obtain a better understanding of the amplitude of the measurement uncertainties, in Fig.\,\ref{fig:axrat_comparison} we compare the axis ratios from the \ellipse\ fits with those derived from the \galfit\ models for the NGVS galaxies. The former measurements correspond to the mean $q$ in the $1 \leq r/r_{e} \leq 2$ range. 
The left and right panels in Fig.\,\ref{fig:axrat_comparison} show the axis ratio difference as a function of the \galfit\  \mue\ and $q$, respectively. 
%The error bars for the difference were derived by adding the corresponding formal uncertainties in quadrature. Note that the formal uncertainties in $q$ from \galfit\ are unrealistically low. 
%The data points in Fig.\,\ref{fig:axrat_comparison} are color-coded according to their $\mu_{e}$ as derived from \galfit, with darker (lighter) symbols corresponding to brighter (fainter) effective surface brightnesses. The solid grey line corresponds to the one-to-one relation, while the dashed lines indicate the rms of the difference, $q_{rms} = 0.03$. This is the typical uncertainty of our individual $q$ measurements, and the figure that will be used in the following analysis.
\new{The distributions of axis ratio differences are similar, both in shape and amplitude, to those in Fig.\,\ref{fig:axrat_err}.
The $q$ values derived from these two completely different methods are remarkably consistent, and especially so for objects with high mean surface brightness. Of course, the disagreement increases at fainter \mue, simply because this parameter is directly related to the mean signal-to-noise ratio across the galaxy body. It is also clear that there is no bias in the determination of axis ratios for either round or elongated galaxies.}

\new{In the analysis that follows we therefore use a \mue-dependent uncertainty for the axis ratio measurements.
At fixed \mue\ the uncertainty follows a normal distribution, with a $1\sigma$ dispersion that includes two components.
The first one is the root-mean-square deviation of the \ellipse\ and \galfit\ measurements, $\sigma_{q} = 0.02$, and represents the minimum uncertainty in the axis ratio estimates. 
This component is added quadratically to a \mue-dependent $\sigma_{q}$ computed through fits to the error distribution obtained from the simulations.
The final $1\sigma$  uncertainty is indicated by the solid lines in Fig.\,\ref{fig:axrat_comparison}, and describes well the distribution of $q$ differences.}
%We additionally color-code the symbols in Fig.\,\ref{fig:axrat_comparison}  according to their \galfit\ $q$, and find that there is no apparent dependence of the uncertainty on their shape. 
It is thus clear that the axis ratios for these faint Virgo galaxies are robustly determined irrespective of both the mean effective surface brightness and the apparent ellipticity of the galaxy.

\subsection{Galaxy detection rate as a function of axis ratio and surface brightness}

\begin{figure}[!t]
\includegraphics[angle=0,width=0.5\textwidth]{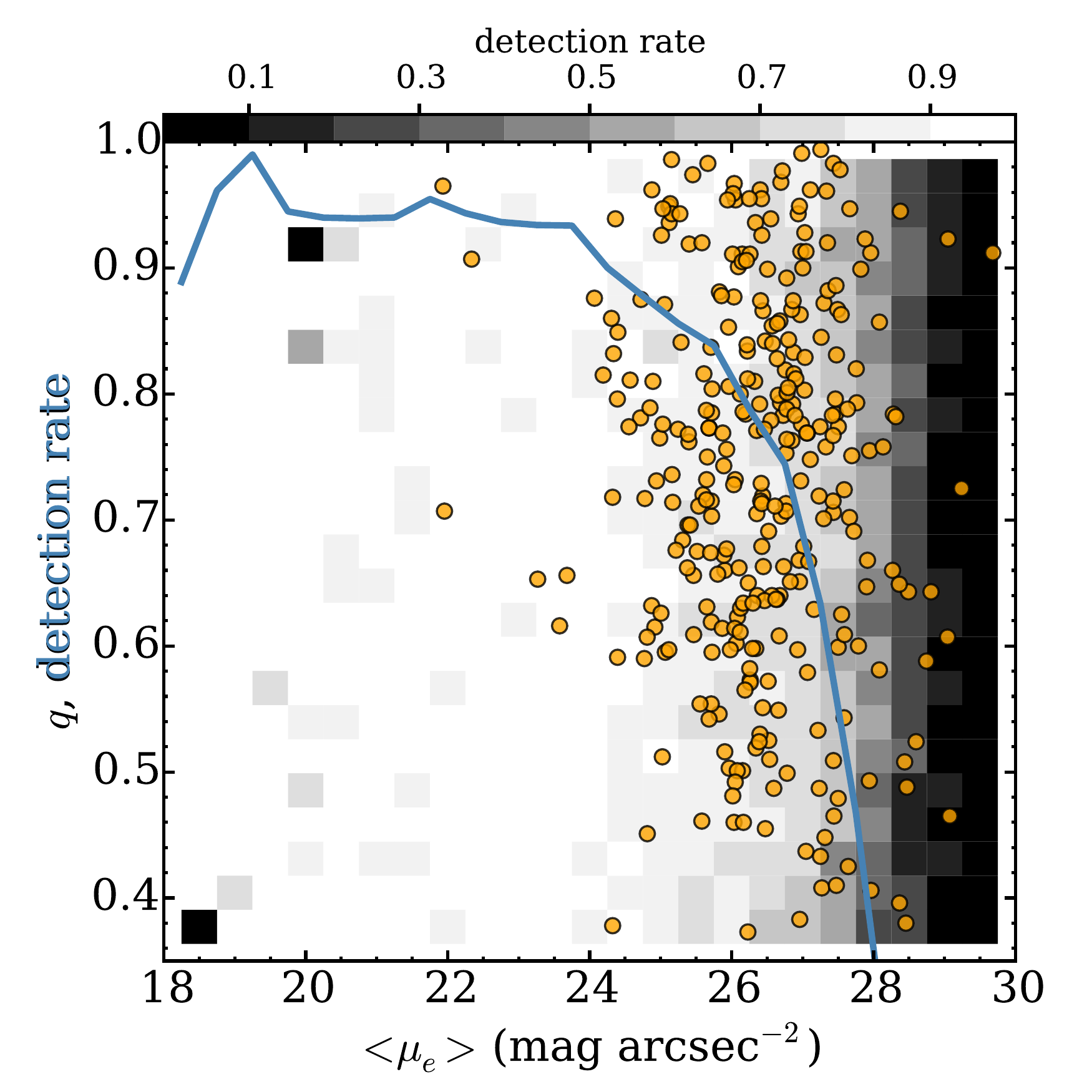}
\caption{\new{Completeness as a function of mean effective surface brightness for simulated S\'ersic galaxies. The grey scale shows the fraction of recovered simulated galaxies as a function of true mean effective surface brightness and true axis ratio. The solid line shows the  detection rate as a function of \mue\ only. NGVS reaches 50 per cent completeness at \mue\ = 27.75 \sb. The symbols correspond to the distribution of real Virgo galaxies.}}
\label{fig:mu_completeness}
\end{figure}

\new{Under the assumption that these low mass galaxies are a family of optically thin triaxial ellipsoids (see Sect.\,\ref{sect:models}), their observed surface brightness at fixed luminosity will depend on the intrinsic triaxiality and the orientation on the sky. 
For example, purely oblate (prolate) systems have brighter surface brightness when seen edge-on (face-on) as a result of larger integration of the emitted light along the line of sight. 
This can introduce a bias against small $q$ values--if the population is oblate--or large $q$ values--if prolate--in the distribution of observed axis ratios at surface brightnesses close to the detection limit of the survey.}

\new{In Figure\,\ref{fig:mu_completeness} we show, for the subsample of red-sequence low-mass galaxies, the distribution of measured axis ratios  as a function of mean effective surface brightness. There is no obvious bias in the distribution of $q$ as a function of \mue. 
The background grey-scale image shows the detection rate for the sample of 36,500 simulated S\'ersic galaxies as a function of these two parameters, and the solid line indicates the completeness as a function of \mue\ only. The few bins showing high incompleteness at high surface brightness levels are caused by a combination of low number statistics, and geometrical non-detections--i.e., objects near the cores of very luminous galaxies, or galaxies that land in the vicinity of bright saturated stars. 
The NGVS reaches 50 per cent completeness at a remarkably faint \mue$_{,lim}$ = 27.75 \sb.
We prefer to be conservative, and in this work we further discard objects with \mue\ values larger than this limit. 
We additionally discard five objects whose proximity to very bright stars or galaxies render their structural parameters highly uncertain.
The final sample of Virgo low-mass galaxies in this study consists of 294 objects. 
We however note that the results of our analysis do not change if we include the fainter \mue\ $>$ \mue$_{,lim}$ objects, which simply is a reflection of the unbiased distribution of $q$ values, and the small subsample size.}
%%%%%%%%%%%%%%%%%%%%%%%%%%

%\section{The radial ellipticity profile of Virgo dwarfs}
%\label{sect:radial}

%The amplitude of the variation for the median profile is 0.18 for bright objects, compared to 0.07 for faint ones.

%\begin{figure}[!t]
%\includegraphics[angle=-90,width=0.5\textwidth]{profiles_transparent.ps}
%\caption{}
%\label{fig:radial_ellip}
%\end{figure}

%%%%%%%%%%%%%%%%%%%%%%%%%%
\section{Bayesian inference of intrinsic shapes}
\label{sect:method}

\subsection{A family of triaxial galaxy models}
\label{sect:models}
The ultimate goal of this study is to derive the underlying distribution of intrinsic shapes for faint galaxies in the Virgo cluster.
While, strictly, full shape information for individual objects can only be derived by a combination of structural and kinematical data \citep[e.g.,][]{Franx1991,statler1994,vandenbosch2009}, it is still possible to constrain the intrinsic shapes of a given population through the observed distribution of apparent axis ratios and a model for their three-dimensional shape. 
In the most general case, galaxies with similar elliptical isophotes are best described by triaxial models \citep{Binney1978}.
One can then try to invert the distribution of apparent axis ratios to infer the distribution of intrinsic shapes \citep{Lambas1992, Lisker2007, Weijmans2014}, but this deprojection problem has no unique solution unless some degree of axisymmetry is assumed--e.g., that the spheroids are either purely oblate, or purely prolate systems \citep{Fall1983}. 
Alternatively, one can try to reproduce the observed distribution of apparent axis ratios by varying the triaxial model parameters, and assuming the galaxies are randomly oriented on the sky \citep{Hubble1926, Sandage1970, Lambas1992, Ryden1994, Sung1998, Vincent2005}. This is the approach we follow in this work.

\begin{figure}
\includegraphics[angle=0,width=.5\textwidth]{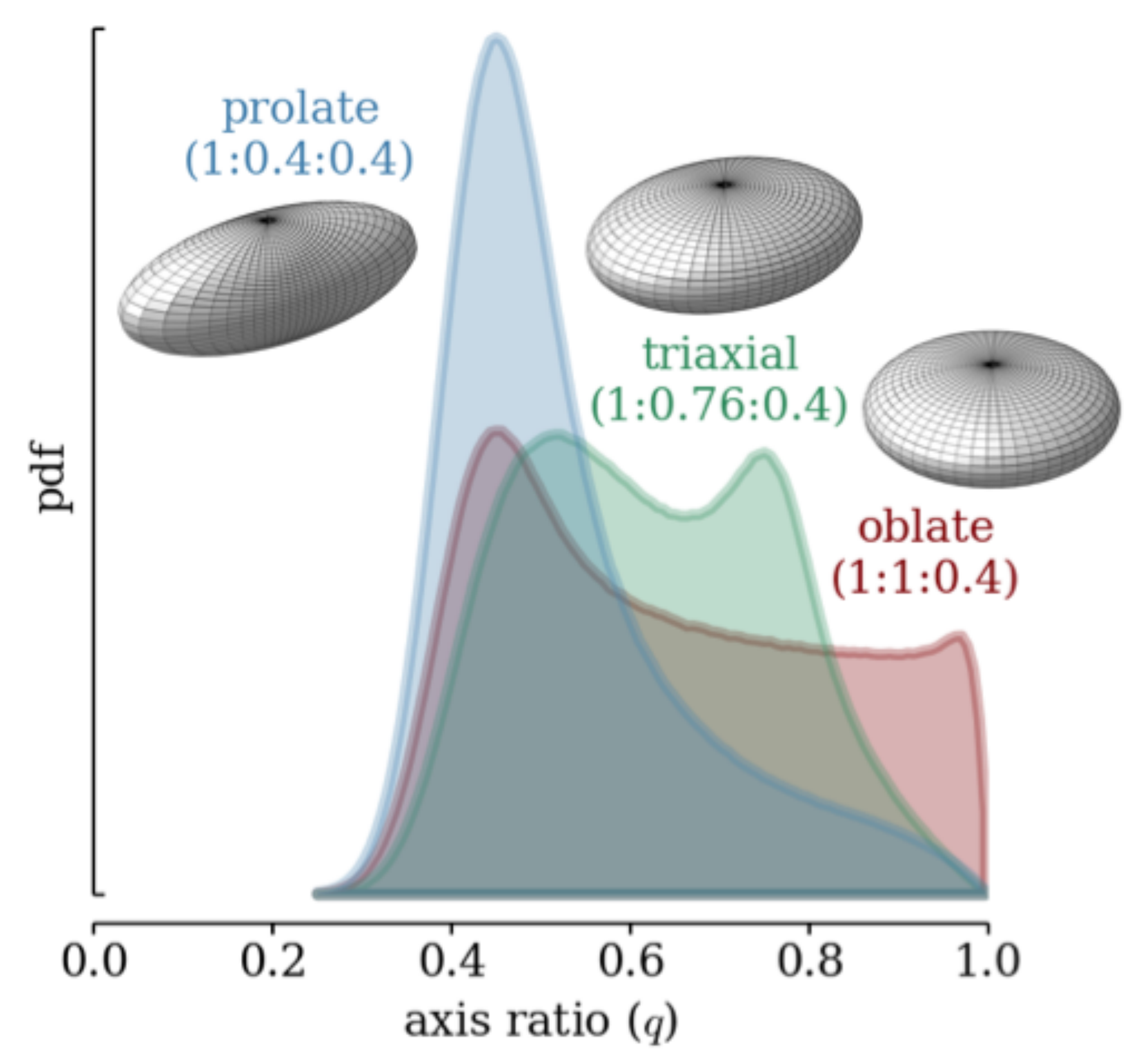}
\caption{Projected axis ratio distributions for three families of triaxial ellipsoids. The populations share a common mean intrinsic ellipticity \ellip $=0.6$, as well as dispersions of ellipticities and triaxialities, namely \sigellip\ = \sigtriax\ = 0.05. The families only differ in the mean triaxiality of the population \triax\ = 0.01, 0.5, and 0.99 for the nearly oblate, triaxial, and nearly prolate populations, respectively. The mean intrinsic  axis ratios $1:B/A:C/A$ for each family are indicated, together with a visualization of the corresponding three-dimensional ellipsoid.}
\label{fig:bapdfs_ex}
\end{figure}

We model faint galaxies in Virgo as a family of optically thin triaxial ellipsoids. 
Following \citet{Franx1991}, we assume that  the three-dimensional galaxy density is structured as a set of coaligned ellipsoids characterized by a common ellipticity $E = 1 - C/A$, and a triaxiality $T = (A^{2} - B^{2})/(A^{2}-C^{2})$, where $A \ge B \ge C$ are the intrinsic major, intermediate, and minor axis of the ellipsoid, respectively.
Special cases of this model include the previously mentioned oblate ($B=A$, or $T=0$) and prolate ($C=B$, or $T=1$) spheroids. More general cases include the oblate-triaxial ($C/B < B/A$) and prolate-triaxial ($C/B > B/A$) spheroids.
In our modeling we further assume that the ellipticity and triaxiality parameters for our galaxy population are normally distributed \citep[see][]{Lambas1992,Ryden1994,Ryden2006,Holden2012}, with means and standard deviations \ellip, \sigellip, \triax, and \sigtriax, respectively. 
Given values for the distributions of intrinsic axis ratios and random viewing angles for each model galaxy, one can easily compute the distribution of apparent axis ratios $q$ \citep{Stark1977,Binney1985}. 

Figure\,\ref{fig:bapdfs_ex} shows three examples of apparent axis ratio distributions constructed from these ellipsoidal models for nearly oblate, nearly prolate, and triaxial families. In this example all models share a common mean intrinsic ellipticity \ellip\ $=0.6$, as well as the same narrow dispersions of ellipticities and triaxialities, namely \sigellip\ = \sigtriax\ = 0.05. The three models only differ in the mean triaxiality of the population, which takes values of \triax\ = 0.01, 0.5, and 0.99 for the oblate, triaxial, and prolate families, respectively.  Fig.\,\ref{fig:bapdfs_ex} illustrates the dramatic impact that this difference makes in the distribution of projected axis ratios, especially at high $q$. While all three populations exhibit a peak  around $q = 1 - \epsilon \sim 0.4$ (associated with the intrinsic mean ellipticity \ellip\ $= 1-C/A =0.6$), the relative number of apparently round objects progressively decreases as a function of \triax. This is because the location of the secondary peak in the observed distribution of $q$ is essentially determined by the mean $B/A$.
A relative paucity of nearly round projected shapes is a generic feature of families of triaxial and prolate ellipsoids.

These apparent axis ratio distributions are equivalent to the likelihood of observing a given axis ratio given the model parameters, and we use this likelihood function to infer the intrinsic shape of low-luminosity galaxies using Bayesian statistics.

\subsection{Bayesian analysis}
\label{sect:bayes}

Given some data $D$ that we want to describe by a model with parameters $\vect{\theta}$, for which we have some prior information $I$, Bayes' theorem states that: 

\begin{equation}
p(\vect{\theta} | D, I) = p(D | \vect{\theta}, I)\,p(\vect{\theta} | I) / p(D | I).
\end{equation}

Here $p(\vect{\theta} | D, I)$ is the posterior probability density function (pdf) of the parameters $\vect{\theta}$ given the data $D$ and prior information $I$; $p(D | \vect{\theta}, I)$ is the likelihood function; $p(\vect{\theta} | I)$  is the prior joint probability of the model parameters in the absence of data; and $p(D | I)$ is the marginal probability, i.e., the a priori probability of having the data marginalized over all possible parameters.
In our particular case, $D$ consists of the observed projected axis ratios for each individual galaxy, $q_{i}$; the model parameters are $\vect{\theta} =$ \{\ellip, \sigellip, \triax, \sigtriax\}; and the likelihood of apparent axis ratios is calculated as explained in Section\,\ref{sect:models}. 
This approach has several formal and practical advantages. 
First, it does not require binning but works directly with discrete data, thus preserving the full information content of the dataset. 
This is always advantageous, but generally not mandatory when the inference is done on large samples of several thousand, or even hundreds of thousand objects \citep[e.g.,][]{Ryden2004,Holden2012}. 
However, the analysis on discrete data becomes really powerful when  the sample size is relatively small. Choosing the optimal binning in these cases is not trivial, it always smears the underlying signal to a certain degree, and statistical uncertainties become non-Gaussian when the bins contain few objects.

In practice, in our Bayesian approach each distribution of axis ratios predicted by the models is convolved with another distribution representing the individual axis ratio $q_{i}$ and its uncertainty. 
This is particularly useful when working with non-uniform and/or non-Gaussian uncertainties, or when the full probability distribution for the measurement of each individual axis ratio is available (e.g., \citealt{Martin2008}).
For the very same reason, it is also straightforward to include the effects of biased measurements, and/or to account for lower or upper limits in the measured axis ratios--e.g. we only know our galaxy has $q > 0.9$.
In this work we model the observed axis ratios as truncated normal distributions with means in the $q \in [0,1]$ interval, and dispersions that are dependent on each galaxy's mean surface brightness, as discussed in Section\,\ref{sect:reliability}. This naturally reproduces the bias towards lower recovered axis ratios at high true $q$ also discussed in that Section. The excess probability from a regular normal distribution at $q > 1$ (which is non-physical) gets redistributed towards $q \leq 1$ values, creating a more extended tail that accounts for the fact that we tend to recover flatter shape measurements due to the bounded nature of axis ratios (see Fig.\,\ref{fig:axrat_err}).

%We use uniform prior distributions for all model parameters in the intervals \ellip\ $\in [0,1]$, \sigellip\ $\in [0,0.5]$, \triax\ $\in [0,1]$, \sigtriax\ $\in [0,0.5]$. Note that, while \ellip\ and \triax\ are by definition bounded, the width of the distributions can in principle be arbitrarily large. However, very large values for these parameters in practice mean that the distribution is unconstrained--any intrinsic ellipticity or triaxiality is possible--, and hence we limit these to exist in the $[0,0.5]$ interval.
\new{We use flat priors for the location parameters of our model normal distributions, i.e., the priors for \ellip\ and \triax\ are uniformly distributed in the interval $[0,1]$.  For the standard deviations \sigellip\  and \sigtriax\ we adopt scale-invariant priors of the form $p(\sigma) \propto \sigma^{-1}$.}
In order to efficiently sample the posterior distribution of our parameters $\vect{\theta}$ we use Markov Chain Monte Carlo (MCMC) methods. In particular, we use the affine-invariant MCMC algorithm implemented in the Python \emcee\ package \citep{ForemanMackey2013}.
We use a set of 100 'walkers' that explore the parameter space at each of the 1000 steps we run. The MCMC chain converges to an equilibrium distribution after approximately $200$ 'burn-in' steps, which we discard when exploring the posterior distributions.
%we use a library of precomputed the projected distributions for individual ellipsoids in a fine grid of $E$ and $T$. The full distribution for the corresponding family of normal models is then constructed through a linear combination of 

%and because the number of model parameters is relatively low, we chose to fully sample the parameter space using precomputed distributions from a grid of $E$ and $T$. 
%%%%%%%%%%%%%%%%%%%%%%%%%%

%%%%%%%%%%%%%%%%%%%%%%%%%%
\section{The intrinsic shapes of  low-luminosity Virgo galaxies}
\label{sect:ngvs_shapes}

\begin{figure*}
\includegraphics[angle=0,width=1.\textwidth]{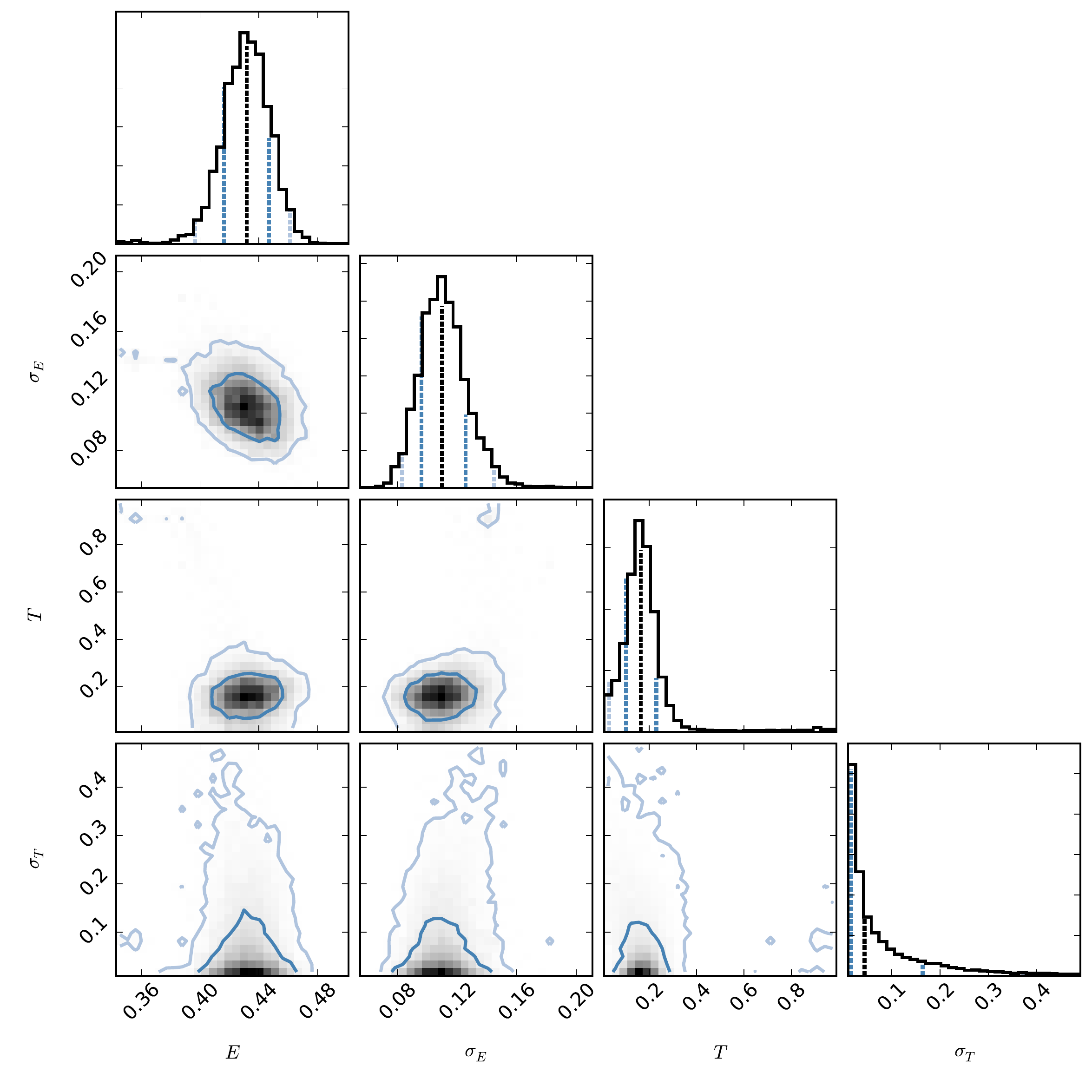}
\caption{Posterior probability density functions for \ellip, \sigellip, \triax, and \sigtriax\ for faint satellites in Virgo. The panels in the diagonal show the posterior pdfs for each of the parameters, marginalized over all the other ones. The grey scale in the non-diagonal panels shows the corresponding joint posterior pdfs. Contours enclose the regions that contain 68 and 95 per cent of the cumulative posterior probability.  The dotted lines in the diagonal panels indicate the 2.5, 16, 50, 84 and 97.5 percentiles of the corresponding marginalized posteriors.}
\label{fig:ngvs_posterior}
\end{figure*}

Figure\,\ref{fig:ngvs_posterior} shows the MCMC posterior distributions for \ellip, \sigellip, \triax, and \sigtriax\ for our sample of faint satellites in Virgo. The panels on the diagonal show the marginalized posteriors for each parameter, and the dotted lines indicate the corresponding 2.5, 16, 50, 84 and 97.5 percentiles.
The non-diagonal panels correspond to the joint posterior pdfs, and contours in these panels enclose the regions that contain 68 and 95 per cent of the cumulative posterior probability.
There are several interesting results worth discussing in detail. 
First, the distribution of intrinsic ellipticities is well constrained by the data, with median and 68 per cent confidence values \ellip $= \Engvs^{\Eungvs}_{\Elngvs}$, and \sigellip $= \SEngvs^{\SEungvs}_{\SElngvs}$. 
The inferred triaxiality distribution of the population is \triax $= \Tngvs^{\Tungvs}_{\Tlngvs}$ and \sigtriax $= \STngvs^{\STungvs}_{\STlngvs}$. It is clear that the triaxiality distribution is not  well constrained with photometric data alone, something known since the early work from \citet{Binggeli1980}.
Nevertheless, the data clearly favor nearly oblate models over nearly prolate ones, although in any case some degree of triaxiality seems to be required. This is best appreciated in the marginalized posterior panel for \triax: the maximum of the distribution is clearly different from zero, but fully triaxial and prolate shapes (\triax\ $\gtrsim 0.4$) are essentially ruled out by the data.

%Moreover, a significant contribution of the likelihood for high-\triax\ values comes from models with very broad triaxiality distributions (see the \triax\ - \sigtriax\ panel), which are not very informative.
%The joint posterior panel for \ellip\ and \triax\ shows that there is a clear anti correlation between these two parameters, in the sense that nearly oblate (prolate) models require a larger (smaller) intrinsic ellipticity. 
%It is important to note that, formally, the marginalized posterior for \sigtriax\ is improper, in the sense that its integral diverges. 
%This is better appreciated in the panel showing the joint posterior distribution for \triax\ and \sigtriax\, where it can be seen that the joint posterior probability is almost flat, but different from zero, for high \sigtriax\ values. 
%The panel showing the joint posterior distribution for \triax\ and \sigtriax\ shows that triaxial and prolate models (\triax\ $\gtrsim 0.5$) are disfavored. %, but only clearly ruled out if the triaxiality distribution for the population is narrow. 
%We note that very large values of either \sigtriax\ or \sigellip\ imply little knowledge about the actual distribution of triaxialities or ellipticities, respectively.
%0.42+/-0.10 for oblate, good fit.
%0.35+-0.12 for prolate, not a good fit

\begin{figure}[!t]
\includegraphics[angle=0,width=0.5\textwidth]{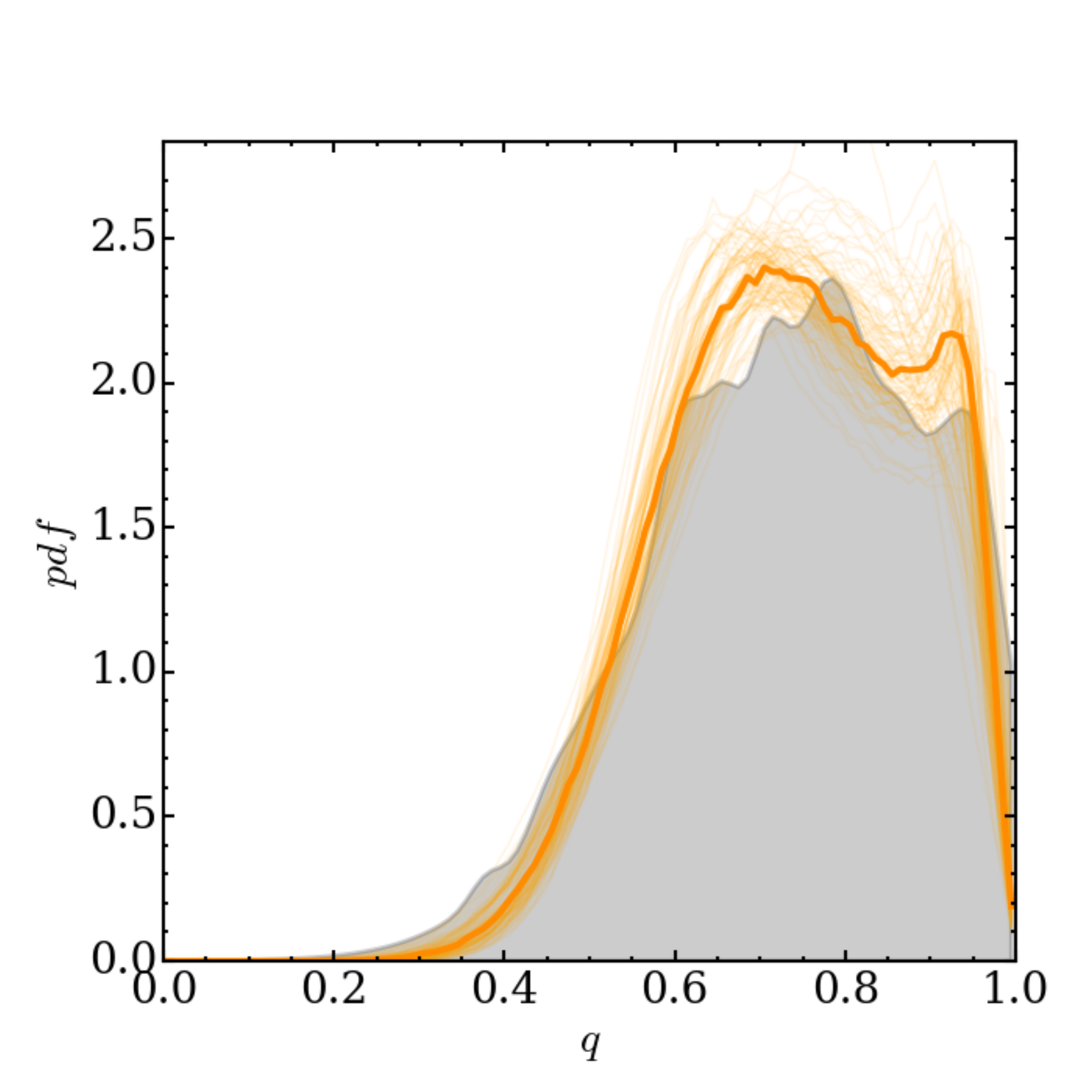}
\caption{The grey area shows a gaussian kernel density estimate of the observed axis ratios of faint galaxies in Virgo. The model corresponding to the median of the samples in the marginalized distributions is shown as a thick solid line, and corresponds to a family of nearly oblate spheroids with mean intrinsic ellipticity \ellip\ $ = \Engvs$. Thin lines correspond to 100 random samples from the MCMC chain.}
\label{fig:ngvs_fit}
\end{figure}

Figure\,\ref{fig:ngvs_fit} shows in grey a gaussian kernel density estimate of the observed axis ratio distribution. Each galaxy is represented by a truncated gaussian located at the corresponding measured $q_{i}$, and a with dispersion determined by its mean surface brightness (see Sect.\,\ref{sect:reliability}). Thin lines correspond to 100 random samples from the MCMC chain, and the thick solid line indicates the model corresponding to the median of the samples in the marginalized distributions.
For reference, the median estimates for \ellip\ and \triax\ correspond to a triaxial ellipsoid with intrinsic axis ratios $1:0.94:0.57$--or, alternatively, it implies that faint Virgo galaxies are nearly oblate, thick spheroids.
%These results imply that Virgo dwarfs are nearly oblate spheroids, with an intrinsic thickness, or minor-to-major axis ratio $C/A \approx 0.57$.

\new{The slight deviation from axisymmetry implied by \triax\ $\neq 0$ has to be taken with some caution. It is mostly driven by the secondary peak at $q \sim 0.94$ in the observed distribution of axis ratios, and by the rapid decline in the number of even rounder galaxies. This relative paucity of round objects has been observed before, notably most recently by \citet{Weijmans2014} in their study of the intrinsic shapes of early-type galaxies in the \atlas\ sample. The positive bias toward lower $q$ introduced by the bounded nature of axis ratio measurements is too small of an effect to explain the lack of round objects in their sample of fast rotators.
Perhaps even more puzzling is the fact that incorporating the kinematic misalignment information in their analysis resulted in a strong support for axisymmetry. Hence \emph{real} triaxiality--understood as $B/A \neq 1$ and a non-axisymmetric orbital structure--may not be the only explanation for the relative paucity of round galaxies. Alternative possibilities include the existence of a small $m=1$ mode, the presence of faint tidal features, or the suggestion that these galaxies may not be fully relaxed.
We lack kinematic information for our NGVS low-mass galaxies, and therefore it is hard to say whether the small degree of triaxiality we infer is real, or related to the previous effect.
In any case, if we impose pure oblateness as a prior (\triax\ = \sigtriax = 0) we infer an intrinsic flattening \ellip $= 0.42 \pm 0.01$ and \sigellip $=0.10 \pm 0.13$, which are identical to the results from the full analysis--but in this case we of course cannot reproduce the observed $q$ distribution at the roundest shapes.}

We finally note that the models we are using implicitly assume that any obscuration by dust is independent of the viewing angle. This is a reasonable assumption, as we expect quiescent galaxies of these low masses to be optically thin systems with negligible amounts of dust \citep{deLooze2010,Boselli2014}. This is true for any low-mass, quiescent galaxy, but more so for these objects in the core of Virgo, where the stripping of the gas and dust components by ram-pressure from the intracluster medium occurs on short timescales \citep[e.g.,][]{Boselli2008}.
%%%%%%%%%%%%%%%%%%%%%%%%%%

%%%%%%%%%%%%%%%%%%%%%%%%%%
\section{Comparison with other galaxy samples and simulations}
\label{sect:discuss}

\subsection{Comparison with luminous early-type galaxies}

\new{The body of literature on the intrinsic shapes of early-type galaxies is enormous, and it is beyond the scope of this paper to present a comprehensive review. We will however discuss two important aspects of these studies: the shape dependence on stellar mass, and the constraints provided by kinematic studies.}

\new{In general, studies that make use of large photometric samples find that early-type galaxies have mean intrinsic minor-to-major axis ratios in the 2:3 to 2:5 range. Differences arise because of selection effects, including the definition of what is exactly an early-type galaxy, as well as the mass range under study \citep{Lambas1992, Vincent2005, Padilla2008}.
\citet{Holden2012} find that at the high mass end (M$_{\star} > 10^{11}$ \msun) intrinsically flattened quiescent galaxies are almost absent--they only constitute $\sim10$ per cent of the population--and that these systems are best described as a family of triaxial spheroids with \triax$=0.45$ and \ellip\ $=0.37$.
This leads them to suggest that M$_{\star} > 10^{11}$ \msun\ sets a mass scale above which disk-destroying multiple major mergers are the
dominant method of mass assembly \citep{vanderWel2009}.
The fraction of oblate spheroids increases toward lower masses, where disks or flattened ellipticals populate the mass function.
Note that pure prolate shapes have long been ruled out for this population of massive early-type galaxies \citep[e.g.,][]{Saglia1993}. }

\new{This result is perhaps best understood by separating early-types into  distinct kinematical classes.
As discussed above, \citet{Weijmans2014} combined photometric and kinematic data for early-type galaxies in the \atlas\ survey to carry out a comprehensive investigation of their intrinsic shapes.
They find that fast rotators (FRs) are much more flattened than slow rotators (SRs). The former population is predominantly axisymmetric, with intrinsic shapes similar to those of spiral galaxies, whereas the latter are best described as mildly triaxial systems. 
The relative abundance of these two families as a function of mass--with SRs dominating at higher masses--naturally explains the prevalence of rounder shapes at the highest masses found by the photometric studies mentioned above.}

\subsection{Comparison with luminous dEs in Virgo}

The first statistical study on the intrinsic flattening of Virgo quiescent galaxies was carried out by \citet{Ichikawa1989} using a sample of nearly one hundred galaxies in the  $-17 < M_{B} < -13$ magnitude range. This early study already showed that Virgo dEs are best described as a population of nearly oblate spheroids with mean intrinsic thickness $\approx\,0.6$.
 \citet{Ryden1994} modeled the intrinsic distribution of axis ratios for a sample of 70 Virgo dEs in the $-17 \lesssim M_{B} < -14$ mag range. They found mean intermediate-to-major and minor-to-major axis ratios of $B/A = 0.95$ and $C/A = 0.54$, respectively. 
 These values are remarkably similar to ours for the fainter population; dEs in their sample also are thick, nearly oblate spheroids. 
%although perhaps these more luminous systems are slightly more flattened than the dwarfs in the Virgo core.
We however note that the subclass of bright nucleated dEs was found to be significantly rounder ($C/A = 0.88$) than the non-nucleated dEs ($C/A = 0.45$; see also \citealt{Ferguson1989}). Here we do not attempt such separation, but it will be explored in a future paper in this series where we extend the analysis to the entire Virgo cluster--hence improving the statistics. 

More recently, \citet{Lisker2007} studied a much larger sample of $\sim 400$ Virgo dEs as faint as $M_{B} = -13$ mag using SDSS imaging data. They inferred the intrinsic axis ratio distribution by inverting the observed $q$ distributions under the assumption that the population is either purely oblate, or purely prolate. They also distinguished between nucleated and non-nucleated systems and, like \citet{Ryden1994} before, found that the latter are significantly flatter than the former. In both cases, though, oblate shapes are favored over prolate ones. For the class of nucleated dEs they found a wide distribution of intrinsic thickness, ranging from 0.5 to 0.8 irrespective of luminosity. They found even more flattened shapes for dEs that show evidence for the presence of spiral arms or residual star formation, suggesting that these systems are probably disky in nature \citep[see also][]{Aguerri2005}.

\new{As discussed above, kinematic data, and especially integral-field observations, are crucial to advance intrinsic shape studies. There is a growing  amount of such data for luminous cluster dEs \citep[e.g.,][]{Toloba2014,Rys2014}, and it would be desirable to augment even more the sample size so that kinematics can be incorporated into the inference process.
This will not only help on constraining axisymmetry, but also in separating the dEs into distinct kinematical classes.
For very faint systems like the ones studied here, however, we will have to rely on building statistically  large photometric samples to constrain their shapes.}

\subsection{Comparison with Local Group satellites}
\label{sect:lg_shapes}

\begin{figure*}
\includegraphics[angle=0,width=1\textwidth]{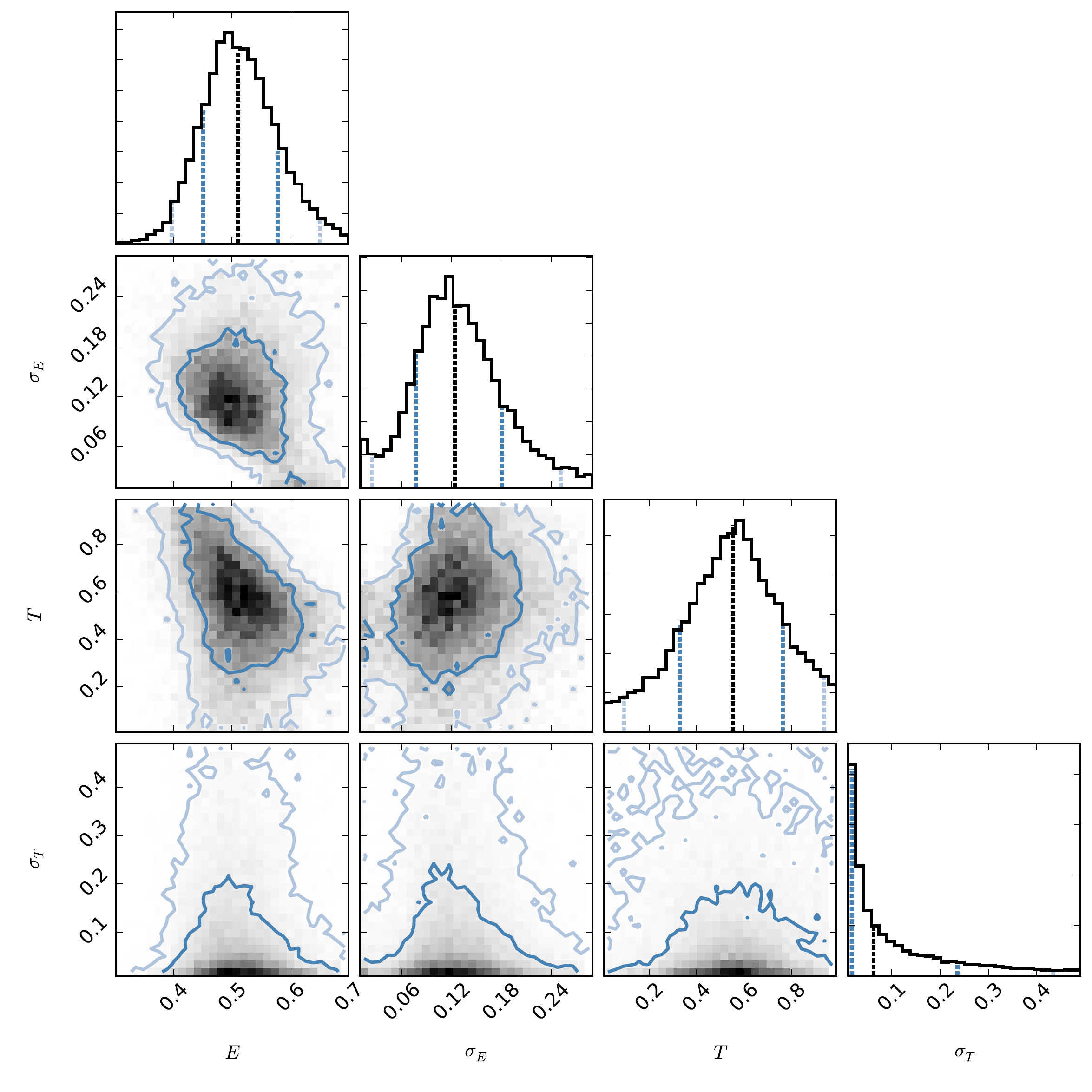}
\caption{Same as Fig.\,\ref{fig:ngvs_posterior}, but for Local Group dSphs. The panels show the posterior probability density functions for \ellip, \sigellip, \triax, and \sigtriax\ for faint satellites in the LG. The panels in the diagonal show the posterior pdfs for each of the parameters, marginalized over all the other ones. The grey scale in the non-diagonal panels shows the corresponding joint posterior pdfs. Contours enclose the regions that contain 68 and 95 per cent of the cumulative posterior probability, respectively.  The dotted lines on the diagonal panels indicate the 2.5, 16, 50, 84 and 97.5 percentiles of the corresponding marginalized posteriors.}
\label{fig:lg_posterior}
\end{figure*}

\begin{figure}
\includegraphics[angle=0,width=0.5\textwidth]{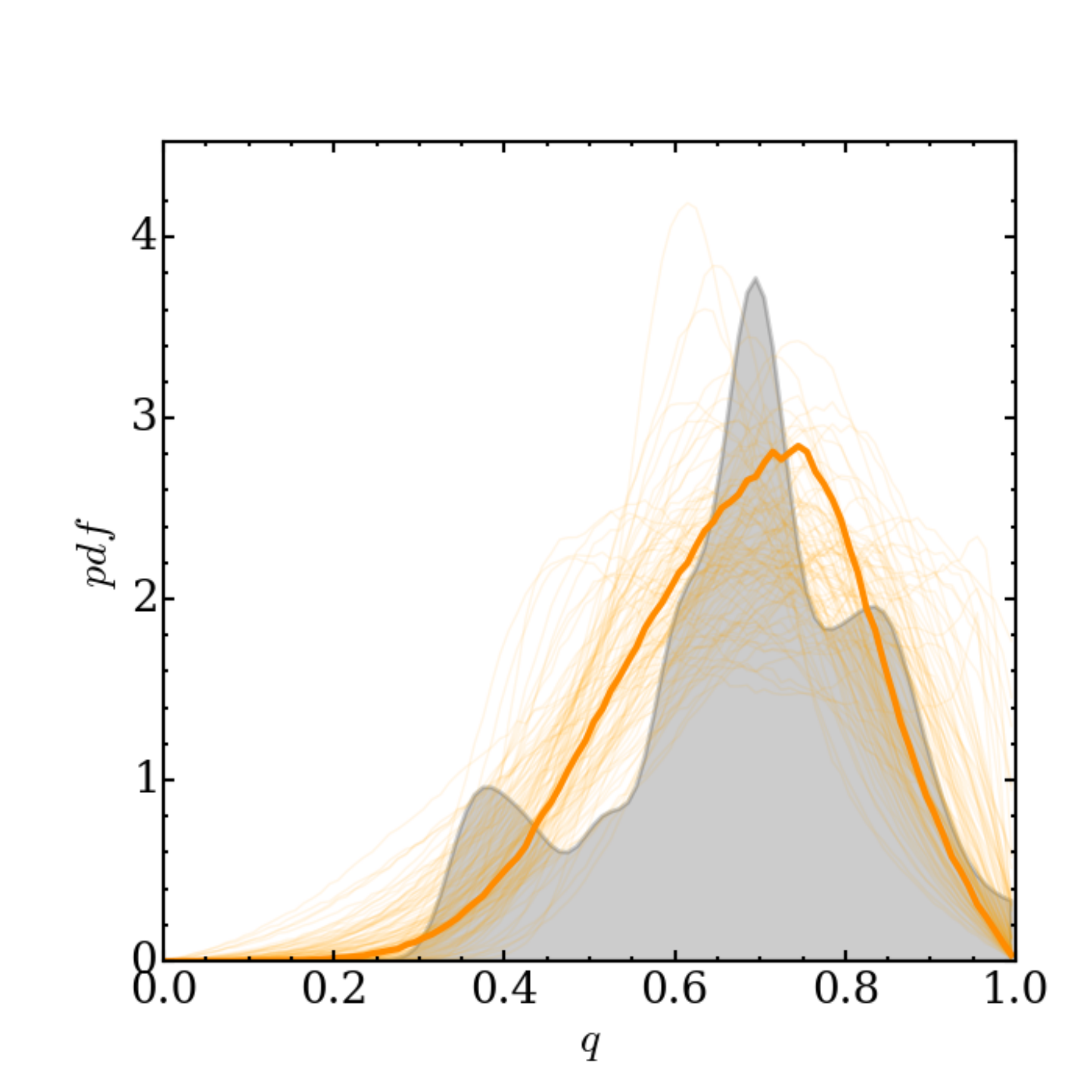}
\caption{Same as Fig.\,\ref{fig:ngvs_fit}, but for Local Group dSphs. The grey area shows a kernel density estimate of the observed axis ratios of faint satellites in the LG. The model corresponding to the median of the samples in the marginalized distributions is shown as a thick solid line, and corresponds to a family of triaxial spheroids with mean intrinsic ellipticity \ellip\ $ = \Elg$. Thin lines correspond to 100 random samples from the MCMC chain.}
\label{fig:lg_fit}
\end{figure}

The Local Group provides a natural comparison sample, because its satellite system comprises objects of similar intrinsic physical properties--e.g., luminosity, gas content, or current star formation rate--to those in Virgo. 
%A more direct comparison with objects of similar intrinsic physical properties (e.g., luminosity, gas content, or current star formation rate) is currently only possible in the Local Group. 
While low number statistics made the task of deriving the intrinsic shapes of LG satellites almost insurmountable in the past, the advent of large panoramic surveys in the last decade has boosted the number of known satellites around both the MW (e.g., \citealt{Koposov2008}), and M31 \citep[e.g.,][]{Richardson2011}.

We carry out an investigation on the intrinsic shapes of LG satellites using the machinery developed in Sect.\,\ref{sect:method}.
We use the structural information for low-luminosity galaxies in the nearby Universe compiled by \citet{McConnachie2012}. In order to select a sample of LG satellites comparable to the ones in Virgo, we introduce a series of cuts. 
First, we limit the sample to actual LG satellites by selecting only low-luminosity galaxies within the zero-velocity surface of the LG (see fig.\,5 and discussion in \citealt{McConnachie2012}). 
Second, we select 'classical' satellites having $-14 \leq M_{V} \leq -8$ mag, thus matching our luminosity range for Virgo systems. Note that restricting the analysis to that luminosity interval is further justified given the claimed discontinuity in apparent dSph shapes below $M_{V} \approx -7.5$ \citep{Martin2008}.
Finally, we discard all systems with M$_{\text{H{\sc i}}}/L_{V} > 0.1~\text{M}_{\odot}/L_{\odot}$. By limiting our sample to objects with low gas mass fractions--or, equivalently, little current star formation activity \citep[e.g.,][]{Weisz2011}--we better mimic our selection of red-sequence faint cluster  galaxies. Satellites with a significant neutral gas component can have different intrinsic shapes than their gas-poor counterparts \citep[e.g.,][]{Roychowdhury2013}, which is natural in a scenario where young stars form within the highest density regions of a kinematically cold H\,{\sc i} disk. 
With this cut we avoid potential biases in our analysis, and it results in a final sample of 23 LG satellites. 
By combining galaxies from both M31 and the MW, we mitigate to some degree the potential complications arising from observing the latter population from a preferred viewing angle \citep{Barber2015}.
The axis ratio measurements for most M31 satellites come from the homogeneous statistical analysis by Martin et al. (2015). For the remaining galaxies, we use the $q$ values and uncertainties as compiled by \citet{McConnachie2012}.
%, but we set the minimum $\sigma_{q}$ value to 0.015--the same as for our Virgo dwarfs, see Fig.\,\ref{fig:axrat_comparison}--for those objects with smaller (most likely unrealistically low) reported uncertainties.  
%The axis ratio measurements for these two galaxies are set exactly to $q=1$ in the original catalogue, but this only corresponds to a rough estimate of their actual ellipticity--and no uncertainties for $q$ are provided. A detailed inspection of the source for these measurements \citep{Ibata2007} reveals that at least AndXV is certainly not perfectly round. We therefore use $q=0.9$ for AndXV and $q=0.99$ for AndXVI, and furthermore assign conservatively large uncertainties $\sigma_{q} = 0.15$ to these two objects. Because of these rather large uncertainties, we have verified that our results do not change significantly if we modify their ellipticities by $\approx 0.05$.
%For these objects we set the minimum $\sigma_{q}$ value to 0.02--the same as for our Virgo dwarfs, see Fig.\,\ref{fig:axrat_comparison}--for those objects with smaller (most likely unrealistically low) reported uncertainties. 

Figure\,\ref{fig:lg_posterior} shows the MCMC posterior distributions for \ellip, \sigellip, \triax, and \sigtriax\ for the final sample of 23 LG satellites.
It is immediately clear that the much smaller sample size translates into poorer constraints on their intrinsic shapes compared to the Virgo sample.
Nevertheless, the intrinsic ellipticity is relatively well constrained, and the mean and dispersion of the population distribution are remarkably similar to those of faint Virgo  galaxies: \ellip\ $= \Elg^{\Eulg}_{\Ellg}$ and \sigellip $= \SElg^{\SEulg}_{\SEllg}$, respectively.
The intrinsic triaxiality of the LG population, however, is significantly different from that of the Virgo sample. LG satellites are best reproduced by triaxial models (\triax\ $= \Tlg^{\Tulg}_{\Tllg}$).
% even though the width of the distribution is largely unconstrained (\sigtriax\ $= \STlg^{\STulg}_{\STllg}$). 
%Note that, as before, \sigtriax\ is an improper parameter, as its posterior diverges. 
%Moreover, the joint \sigtriax-\triax\ posterior panel in Fig.\,\ref{fig:lg_posterior} shows that triaxial models are preferred over nearly oblate and prolate ones if the dispersion is low--that is, if the triaxiality distribution of the population is not too wide.
%0.49 +/- 0.11, or C/A=0.50. Terrible fit
%0.41 +/- 0.13, or C/A=0.59. Very decent fit.

The difference in mean triaxiality between the LG and Virgo low-mass galaxies can be readily understood with the help of Fig.\,\ref{fig:lg_fit}. As before, the grey area shows a kde estimate of the observed axis ratios in the LG. Thin lines again correspond to 100 random samples from the MCMC chain, and the thick solid line indicates the model corresponding to the median of the samples in the marginalized distributions. 
The relevant feature of this figure is the marked paucity of objects with very round apparent shapes. As discussed in Sect.\,\ref{sect:models} and illustrated in Fig.\,\ref{fig:bapdfs_ex}, this is a characteristic of intrinsically triaxial or prolate populations, simply because random projections on the sky of these objects rarely result in perfectly circular shapes.

The inferred mean values for the model distributions correspond to a triaxial ellipsoid with intrinsic axis ratios $1:0.76:0.49$.
\new{We note that purely oblate shapes (\triax = 0) provide a poor description of the observed axis ratio distribution. Purely prolate shapes (\triax =1), on the other hand, do a better job at reproducing the distribution in Fig.\,\ref{fig:lg_fit}, but result in thicker spheroids, \ellip $= 0.41 \pm 0.04$ and \sigellip $= 0.13^{+0.04}_{-0.03}$. This is already evident in the \ellip-\triax\ panel of Fig.\,\ref{fig:lg_posterior}, where the contours enclosing 95 per cent of the posterior probability extend toward less flattened shapes at high \triax.}

%To our best knowledge, no previous work has attempted an investigation of the intrinsic shapes of LG dwarfs, with studies so far focusing on the distribution of apparent ellipticities.
\citet{Martin2008} presented a comprehensive analysis of the structural properties of faint satellites around the MW, including their \emph{apparent} ellipticities. 
They found that objects in the $-14 \lesssim M_{V} \lesssim -7.5$ magnitude range (essentially the very same ones used here) have relatively round shapes, with a mean observed  $q \approx 0.7$ (compare with the peak of the observed distribution in Fig.\,\ref{fig:lg_fit}). Remarkably, there seems to be a transition in galaxy shape around this luminosity, as fainter systems have more elongated shapes (mean $q \sim0.5$), and a broader distribution of apparent axis ratios. They explore several possibilities for the origin of these very elongated shapes, including flattening by rotation, a more elongated shape of their dark matter halos, and tidal distortions. They argue that the latter scenario--that tidal interactions with the MW drive dSph shapes--is slightly preferred over the other mechanisms,  and suggest that the broad distribution of observed ellipticities ($ 0.2 \lesssim \epsilon \lesssim 0.8$) may correspond to different stages in the disruption process (cf. Sect.\,\ref{sect:sims}). 

\new{Recently, \citet{Salomon2015} have inferred the intrinsic ellipticity of M31 satellites under the assumption that their visible components have prolate shapes, and after exploring different possibilities for the orientation of the galactic major axis. They find that the population has a mean intrinsic axis ratio $C/A \sim 0.5$, which is virtually identical to the value derived here.  A non-negligible fraction of the studied satellites (6/25) are found to be intrinsically significantly elongated, and these authors suggest they may be strongly affected by tides.
This, in turn, would have important implications for the orbital distribution of M31 dSphs, because significant tidal stretching is known to be a short-lived phase mostly limited to pericentric passages \citep{Barber2015}.}

It would be extremely interesting to test the disruption hypothesis in the case of extremely faint Virgo galaxies because they reside in a much more dense environment than the MW, and have done so for longer times. 
Recently, \citet{Jang2014} reported the serendipitous discovery of a candidate ultra-faint ($M_{V} \approx -6.5$) galaxy in very deep HST/ACS images of the Virgo cluster. The fact that the apparent ellipticity of this one object is much lower ($\epsilon \approx 0.1$) than what is generally found in the LG opens up the exciting possibility that these ultra-faint systems have different intrinsic shapes in groups and clusters.
This is admittedly pure speculation, as one single object hardly constrains the intrinsic shape of a population. But, unfortunately, relatively large samples of such galaxies in this very low-mass regime will remain inaccessible in Virgo for years to come. 

%It is however interesting that down to $M_{V} \sim -7.5$ mag both the Virgo and MW dSphs have very similar mean ellipticities. 
%This might indicate that in this mass regime the depth of the host potential well is not driving the structural properties of these galaxies, and opens up the possibility that the intrinsic shape is instead governed by the stellar mass of the galaxy. 

\subsection{Comparison with numerical simulations of tidal evolution for satellites}
\label{sect:sims}

Tidal interactions between a satellite and its host, as well as with other subhalos, are essentially inevitable during their orbital evolution. 
Numerical simulations show that tidal stripping and shocking generally result in significant mass loss from the dark matter halo \citep[e.g.][]{Gnedin2003,Penarrubia2008,Smith2013}. Depending on the strength and duration of the interaction, and on the properties of the progenitor galaxy, tidal effects can also drive significant morphological transformation, redistribution of the stellar orbits, and even inflict stellar mass losses. 

\new{\citet{Barber2015} studied the shapes of the gravitational potential of subhalos in the Aquarius simulations at radii comparable to the typical half-light radii of LG dSphs. They show that inner isopotentials are nearly prolate, with intrinsic axis ratios $B/A \approx 0.8$ and $C/A \approx 0.75$. While departure from sphericity is in good agreement with our results for the LG satellites, the mean thickness inferred from observations, $C/A \approx 0.5$, is significantly lower than that of simulated subhalos \citep[see also][]{Salomon2015}. The cumulative effect of continuous tidal stripping, however, is to reduce the asphericity of the subhalos.
It is tempting to associate this effect to the rounder shapes we have inferred for Virgo low-luminosity galaxies compared to LG dSphs. Not only is tidal stripping stronger in the cluster potential, but a large fraction of cluster galaxies have spent many Gyr orbiting within group-sized subhalos before being accreted onto the cluster \citep{Lisker2013}.
For this to be the case, however, requires the host subhalo of these galaxies to be heavily stripped ($>85-90\%$ of the initial dark matter mass; see \citealt{Smith2013,Smith2015,Mistani2016}).
Strong tidal stripping can occur as the satellite galaxies orbit a more massive host during their infall onto the cluster \citep[e.g.,][]{Paudel2013}, or by the cluster potential itself.
 In fact, the  field under study contains some obvious cases of tidally stripped  low-luminosity galaxies--like VCC1297, which is a prototype for the more luminous cE class \citep[see also][]{Guerou2015}.}
 
An even more dramatic transformation, in terms of both mass loss and redistribution of stellar orbits, can occur if the progenitor galaxy is a dynamically cold disk.
In this case, the cumulative effects of tidal stripping, tidal shocking, and resonant stripping result in the creation of a lower mass, spheroidal-like system \citep{Moore1996, Moore1998, Mayer2001, Mastropietro2005, DOnghia2009, Smith2010, Kazantzidis2011}. 
Naturally, these mechanisms have been invoked to account for the morphology-density relation in both galaxy-sized and cluster-sized halos.

\begin{figure}[!t]
\includegraphics[angle=0,width=0.5\textwidth]{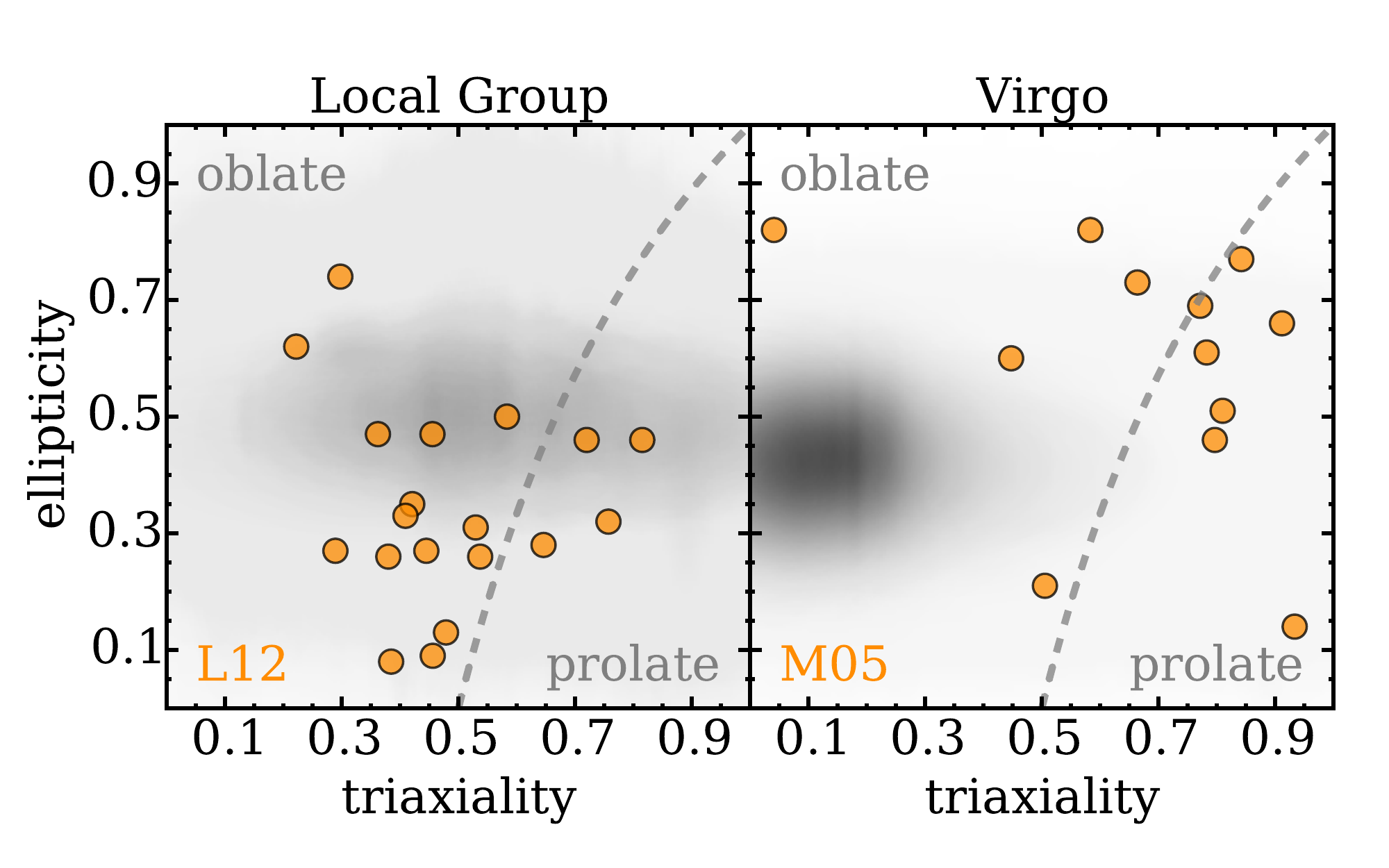}
\caption{Comparison between the inferred intrinsic ellipticity and triaxiality for faint galaxies in the Virgo cluster and in the Local Group, and the predictions from numerical simulations of tidally modified disks orbiting cluster- and galaxy-sized halos, respectively. In both panels, the shaded region corresponds to 1000 random bivariate normals from the MCMC chain for the model parameters. The dashed line marks the transition between prolate-triaxial and oblate-triaxial spheroids. In the left panel, the symbols show the results for the set of simulations of low-mass disks evolved in a MW-like potential by \citet{Lokas2012}. In the right panel we show the final shapes of discs evolved within a Virgo-like halo \citep{Mastropietro2005}. Note that, while the LG simulations roughly match the inferred distribution of intrinsic shapes, the remnants simulated in the cluster environment are both too flattened and too prolate compared to observed faint Virgo galaxies (but see text for discussion).}
\label{fig:sims}
\end{figure}

In Figure\,\ref{fig:sims} we compare the intrinsic shapes of faint galaxies in the LG and Virgo inferred in the present work with the predictions from two sets of numerical simulations of tidally transformed disks.
For the comparison with the LG satellites we use the set of simulations from \citet{Lokas2012}, who follow 10 Gyr of orbital evolution of a galaxy with initial stellar mass $\text{M}_{\star} \approx 2\times10^{7}~\text{M}_{\odot}$ within an MW-sized halo. The final intrinsic ellipticity and triaxiality of the remnants is shown with round symbols in Fig.\,\ref{fig:sims} (left panel). The shaded region corresponds to 1000 random bivariate normals from the MCMC chain for the model parameters of LG satellites (Sect.\,\ref{sect:lg_shapes}). The dashed line marks the transition between prolate-triaxial and oblate-triaxial spheroids.
There is reasonably good agreement between the data and the simulations, with the majority of the simulated remnants being oblate-triaxial spheroids with moderate to small intrinsic ellipticities. 
The three simulated galaxies with the roundest shapes--which are the most tidally stripped systems--are the least consistent with the distribution of intrinsic  shapes inferred in this work.
%There is however a fraction of the simulated dwarfs that are too round compared to the inferred shapes for LG dwarfs. 
It is however important to note that, in their set of simulations, \citet{Lokas2012} vary not only the orbital parameters of the satellite, but also their overall stellar and dark matter halo properties (including the initial stellar thickness and mass, and the dark halo concentration, spin, and mass). Their simulations can only be interpreted as being representative of the range of shapes one can expect from the evolution under the tidal stirring scenario, and never as predicting a full distribution of realistic shapes.
As discussed before, the paucity of observed LG satellites with very round apparent shapes strongly suggests that the population must have a certain degree of intrinsic triaxiality.

We compare the intrinsic shapes of low-luminosity Virgo galaxies with the set of simulations from \citet{Mastropietro2005}. These authors follow the evolution of a Large Magellanic Cloud-type disk galaxy infalling into a Virgo-sized halo at $z=0.5$. Their simulated galaxies undergo significant mass loss and morphological evolution, and their final intrinsic shapes are shown in the right panel of Fig.\,\ref{fig:sims}.
It is clear that the remnants from this set of simulations do not resemble the population of faint galaxies in the core of Virgo: they are both too flattened and too prolate compared to the inferred intrinsic shapes of low-luminosity galaxies in the NGVS.
We note that in Fig.\,\ref{fig:sims} we only include the simulated remnants that by $z=0$ have experienced at least one orbit with a pericentre bringing them within the virial radius.

However, this striking difference is probably just a limitation of using this particular set of simulations for the comparison.
The galaxies simulated by M05 are initially too massive to be the progenitors of the faint Virgo galaxies in our sample: even their most heavily stripped remnant has a final stellar mass at $z=0$ roughly seven times larger than our most massive object.
%Second, because of its relatively high mass for a dwarf (M$_{\star} \approx 4\times10^{9}$ M$_{\odot}$), their initial model galaxy is extremely cold from a dynamical point of view--which in turn translates into a very thin initial disk configuration.
This relatively high initial mass (M$_{\star} \approx 4\times10^{9}$ M$_{\odot}$) also implies a high disk self gravity, which coupled to a very thin and cold initial configuration makes the galaxy rather bar-unstable.
Indeed, the severe morphological transformation experienced by the galaxy in these simulations is essentially driven by the formation of a strong bar in the early stages of its tidal evolution. 
The low surface density material remaining around the bar is subsequently easily stripped by tidal forces, leaving behind a prolate-triaxial stellar body (essentially, a naked bar). The bar usually buckles, and the system becomes progressively rounder as additional tidal  interactions and heating occur.

It is important to note that the model galaxy in M05 is so bar-unstable that \emph{all} of their twenty simulated galaxies form a bar at some point during their evolution--including all of the less-perturbed galaxies that do not even penetrate the cluster virial radius during their orbital evolution (see their table 1, and the discussion in section 3 of that paper). 
This is probably unrealistic. While it is true that the bar fraction \emph{in the field} peaks at about 50 per cent around a stellar mass M$_{\star} \sim 10^{9}$ M$_{\odot}$, only one out of five Virgo disk galaxies are barred in this stellar mass regime--and the bar fraction quickly drops to zero at even lower masses \citep{Mendez-Abreu2010,Mendez-Abreu2012}. This suggests that bar-driven evolution due to tidal interactions in the cluster environment is less important than these simulations predict.  This scenario is fully consistent with the lack of prolate, bar-like satellites in Virgo. 

It is possible that the apparent discrepancy between the inferred intrinsic shapes of faint Virgo galaxies and those predicted by numerical simulations of tidal interactions can be reconciled simply by simulating more adequate models for their progenitor galaxies. For instance, it is clear now that 'central' (as opposed to satellite) galaxies with masses  M$_{\star} < 10^{9}$ M$_{\odot}$ are not thin disks, but rather thick oblate spheroids \citep{rsj2010, Roychowdhury2013} in which pressure support is dominant \citep{Wheeler2015}. Additionally, \citet{Kazantzidis2011} show that tidal shocks heat the stellar body more effectively in lower mass systems, and as a result the remnants evolve into more spherical shapes.
\new{Note, however, that these simulations do not succeed in fully removing rotation, but rather decrease $v/\sigma$ by increasing the stellar velocity dispersion. Therefore, they do not provide a perfect description of the faintest LG satellites--which lack rotation support--, but transformation through tidal stirring may be an appropriate scenario for more massive dE-type galaxies.}
In any case, and to our best knowledge, no simulations have yet studied the tidal evolution of galaxies in the low luminosity regime explored in this work within a Virgo-like environment.

\subsection{Comparison with low-luminosity, star-forming galaxies in the field}

Finally, our results can also be compared with the intrinsic shape of faint star-forming galaxies, i.e., similar mass objects residing in lower density environments and having later-type morphologies and measurable star formation rates. \citet{Sung1998} already noted that faint star-forming galaxies do not conform to the belief that they are highly flattened systems; much on the contrary, they actually are relatively puffy objects. \citet{rsj2010} later showed that, under the assumption that low-mass galaxies are well described as oblate spheroids, their thickness systematically increases towards fainter luminosities. They argued this trend to be an intrinsic property--faint galaxies in their sample are mostly centrals, not satellites--related to the complex interplay between mass, specific angular momentum, and stellar feedback effects. Simply put, the increasing importance of turbulent motions over rotational support in lower mass galaxies leads to the formation of thicker systems \citep[e.g.][]{Kaufmann2007}.
 The recent kinematical analysis of low-mass galaxies in the Local Volume by \citet{Wheeler2015} provides strong support for this scenario. These authors find that the great majority of M$_{\star} < 10^{8}$ M$_{\rm \odot}$ galaxies, regardless of whether they are isolated or satellites, are pression-supported.

A more quantitative comparison with the results presented here can be made by using the recent analysis of \citet{Roychowdhury2013}.
They follow a similar approach to ours to derive the intrinsic shapes of a sample of several hundred star-forming faint galaxies in the Local Volume. 
\new{Their optical axial ratios were measured at the Holmberg isophote, $\mu_{B} = 26.5$ \sb, and consequently they should be a fair representation of the underlying galaxy shape rather than being affected by recent star formation events.}
These authors find the same trend as \citet{rsj2010}, namely that fainter star-forming systems are systematically rounder. Their lowest-luminosity subsample ($-13 \lesssim M_{B} \lesssim -7$ mag) is a reasonably good match to the luminosities of the faint Virgo systems studied in this work. These nearby dIrrs are best described by a family of oblate spheroids with mean intrinsic ellipticity \ellip\ $ = 0.52$.
This is again a very similar value--though implying slightly flatter shapes--to the one we derive here for both Virgo and, especially, LG quiescent galaxies of the same luminosity. 
This finding strongly indicates that, to first order, the mechanisms shaping the structure of these low-mass systems are almost independent of the \emph{present} local environment. 
 If the progenitors of low-luminosity Virgo galaxies did have shapes similar to present-day faint dIrrs--an assumption that does not necessarily have to hold--then the environmental effects that transformed them into non-star forming galaxies have resulted in a relatively small amount of thickening of their stellar bodies.
%%%%%%%%%%%%%%%%%%%%%%%%%%

%%%%%%%%%%%%%%%%%%%%%%%%%%
\section{Summary and conclusions}
\label{sect:summary}

We have investigated the intrinsic shapes of faint, red-sequence galaxies in the central 300 kpc of the Virgo cluster using deep imaging obtained as part of the NGVS.
We have built a sample of nearly 300 red-sequence cluster members in the $-14 < M_{g} < -8$ magnitude range, and we measure their apparent axis ratios through S\'ersic fits to their two-dimensional light distribution in the $g$-band.
The resulting distribution of apparent axis ratios is then fit by a family of triaxial models with normally-distributed intrinsic ellipticities and triaxialities. 
We develop a Bayesian framework to explore the posterior distribution of the model parameters. This approach allows us to work directly on discrete data (i.e., without binning), as well as to account for individual, surface brightness-dependent axis ratio uncertainties. 

We find that faint galaxies in the core of Virgo are best described as thick, nearly oblate spheroids, with mean intrinsic axis ratios $1:0.94:0.57$.
Using the machinery developed for this investigation, we additionally carry out a study of the intrinsic shapes of satellites of similar luminosities in the Local Group.
For this population we infer a slightly larger mean intrinsic ellipticity, but the lack of objects with round appearances translates into more triaxial shapes, $1:0.76:0.49$.
We compare these results with numerical simulations that follow the tidal evolution of satellites within cluster- and group-sized halos.
The LG simulations generally match the inferred distribution of dSph intrinsic shapes, but the remnants simulated in the cluster environment are both too flattened and too prolate compared to observed faint Virgo galaxies. 
We discuss possible reasons for this discrepancy, and conclude that more adequate simulations of cluster galaxies in the luminosity regime explored in this work are needed in order to compare with the observations.

We finally compare the intrinsic shapes of the NGVS low-luminosity galaxies with samples of more massive, quiescent cluster systems, and with star-forming galaxies of similar luminosities residing in lower-density environments. Remarkably, all these low-mass galaxy samples are also best described as thick, nearly oblate spheroids, and their intrinsic ellipticities and triaxialities are consistent, within the uncertainties, with those inferred for the faint Virgo population. 
This is an important result, as it implies that, to first order, \emph{intrinsic shapes in this luminosity regime are largely independent of the environment in which the galaxy resides.}

There is, however, a hint that cluster objects may be slightly thicker than group satellites, which are not different from more isolated faint galaxies. Indeed, the mean thickness of the low-luminosity population is $C/A = 0.57, 0.49,$ and $0.48$ for objects in Virgo, in the LG, and in the field, respectively.
The fact that low-mass galaxies that have a priori experienced very different degrees of environmental effects have very similar flattenings seems to suggest that internal--and not external--mechanisms are the main drivers of galaxy structure in this luminosity regime. 
This may be the case if, for example, galaxies of these masses already form as thick, puffy systems due to baryonic processes that act to pressurize gas within low-mass halos. These processes may include photoionization from the cosmic UV background, and star formation feedback \citep{Kaufmann2007, Teyssier2013}.  Alternatively, other sources of internal heating, such as dynamical interactions with dark subhalos may play a role as well \citep{Starkenburg2015}.
Subsequent gas stripping and tidal evolution within their host halos--especially if acting early on--certainly can contribute to some degree to the heating of  the stellar body of satellites, making them progressively thicker.
However, we cannot rule out the possibility that these systems were already formed with the observed distributions of intrinsic flattenings.

We will explore this scenario in more detail in a future paper of this series using an extended sample of several thousand faint NGVS galaxies across the entire Virgo cluster, for which we are currently deriving structural parameters and establishing robust membership. With increased statistics we will be able to investigate intrinsic shapes as a function of mass, color, and location in position-velocity phase-space, thus addressing the interplay between internal and external processes in shaping the structure of low-luminosity galaxies.

%%%%%%%%%%%%%%%%%%%%%%%%%%

\acknowledgments
We thank an anonymous referee for useful suggestions that have improved the readability of the manuscript. 
RSJ thanks Francis-Yan Cyr-Racine, Seb Fabbro, and Ariane Lan\c{c}on for interesting discussions on Bayesian inference, and Rory Smith for comments on the manuscript. 
We also thank Nicolas Martin for providing us with updated structural parameters for the M31 satellites prior to publication.

This work was supported in part by the National Science Foundation under Grant No. PHYS-1066293 and the hospitality of the Aspen Center for Physics.
S.M. acknowledges financial support from the Institut Universitaire de France (IUF), of which she is senior member.
THP acknowledges support through a FONDECYT Regular Project Grant (No. 1121005) and BASAL Center for Astrophysics and Associated Technologies (PFB-06).

Based on observations obtained with MegaPrime/MegaCam, a joint project of CFHT and CEA/IRFU, at the Canada-France-Hawaii Telescope (CFHT) which is operated by the National Research Council (NRC) of Canada, the Institut National des Science de l'Univers of the Centre National de la Recherche Scientifique (CNRS) of France, and the University of Hawaii.
The authors wish to recognize and acknowledge the very significant cultural role and reverence that the summit of Mauna Kea has always had within the indigenous Hawaiian community.  We are most fortunate to have the opportunity to conduct observations from this mountain.

This work is supported in part by the Canadian Advanced Network for Astronomical Research (CANFAR) which has been made possible by funding from CANARIE under the Network-Enabled Platforms program. This research used the facilities of the Canadian Astronomy Data Centre operated by the National Research Council of Canada with the support of the Canadian Space Agency. The authors further acknowledge use of the NASA/IPAC Extragalactic Database (NED) which is operated by the Jet Propulsion Laboratory, California Institute
of Technology, under contract with the National Aeronautics and Space Administration, and the HyperLeda database (http://leda.univ-lyon1.fr).

%{\it Facilities:} \facility{Nickel}, \facility{HST (STIS)}, \facility{CXO (ASIS)}.

\bibliography{/Users/sanchezjr/Dropbox/rsj_references.bib}

%\appendix{A note on the alignments of faint dwarfs in the core of Virgo}

\end{document}